\RequirePackage{lineno}
\documentclass[12pt]{article}

\usepackage[disable]{todonotes} 
\usepackage[authoryear, round,longnamesfirst]{natbib}

\usepackage{setspace}
\usepackage{amsmath}
\usepackage{mathtools}
\usepackage{amsthm} 
\usepackage{amssymb}
\usepackage{comment}
\usepackage{lscape}

\usepackage{newtxtext,newtxmath} 

\usepackage{tikz}
\usetikzlibrary{positioning}
\usepackage{color}
\usepackage{xcolor}
\usepackage{graphicx} 
\usepackage{booktabs} 
\usepackage{amssymb}
\usepackage{longtable,booktabs} 
\usepackage{csquotes}
\usepackage{pgfplots}
\usepackage{enumitem}
\usepackage{psfrag} 
\usepackage{epstopdf}
\usepackage{subcaption}
\usepackage{ctable}
\usepackage[flushleft]{threeparttable}

\usepackage{microtype}
\usepackage{wrapfig}
\usepackage[toc,page]{appendix}
\usepackage{tablefootnote}

\usepackage{algcompatible}
\usepackage{algorithm}
\usepackage[noend]{algpseudocode}
\usepackage{matlab-prettifier}
\makeatletter
\def\BState{\State\hskip-\ALG@thistlm}
\usepackage{chngcntr}
\usepackage{arydshln}

\theoremstyle{plain}

\newtheorem{proposition}{Proposition}

\usepackage[margin=1in]{geometry}
\usepackage{color}   
\usepackage{hyperref}
\hypersetup{
	colorlinks=true, 
	linktoc=all,     
	linkcolor=LFBlue,  
	citecolor = LFBlue,
	urlcolor  = LFBlue,
}
	
	\definecolor{LFBlue}{RGB}{0,0,90} 
	\definecolor{OxfordBlue2}{RGB}{4,30,100} 
	\definecolor{Oxfordgrey}{rgb}{0.3686, 0.5255, 0.6235} 
	\definecolor{LFGrey}{rgb}{0.3686, 0.5255, 0.6235} 
	\definecolor{LFOrange}{rgb}{1, 0.5608, 0}
	\definecolor{DarkOrange}{RGB}{114,47,55}
	
\pgfplotsset{yticklabel style={text width=3em,align=right}}

\newcommand\blfootnotea[1]{%
	\begingroup
	\renewcommand\thefootnote{$*$}\footnote{#1}%
	\addtocounter{footnote}{-1}%
	\endgroup
}
\newcommand\blfootnoteb[1]{%
	\begingroup
	\renewcommand\thefootnote{$\dag$}\footnote{#1}%
	\addtocounter{footnote}{-1}%
	\endgroup
}
\newcommand\blfootnotec[1]{%
	\begingroup
	\renewcommand\thefootnote{$\ddagger$}\footnote{#1}%
	\addtocounter{footnote}{-1}%
	\endgroup
}

\usepackage{xr}
\begin{document}


\thispagestyle{empty}
\vspace*{-1cm}
\begin{center}
	{\Large  Monetary-Fiscal Interaction and the Liquidity of Government Debt\blfootnotea{We thank Evi Pappa and two anonymous referees as well as Alice Albonico, Florin Bilbiie, Francesco Caselli, Andrea Colciago, Davide Furceri, Wouter den Haan, George Kouretas, Ben Moll, Ricardo Reis, Silvana Tenreyro and seminar participants at the Bank of England and the LSE for comments and suggestions. Part of this research was conducted at the Bank of England which we thank for the support.
	}} \\
	\vspace*{0.5cm}
	\centering{
		{\large Cristiano Cantore}\blfootnoteb{Sapienza University of Rome. Email: \href{mailto:cristiano.cantore@uniroma1.it }{cristiano.cantore@uniroma1.it }. Web: \href{https://www.cristianocantore.com/}{cristianocantore.com}. }
	}
	\hspace*{1cm}
	\centering{
		{\large Edoardo Leonardi}\blfootnotec{London School of Economics and Political Science. Email: \href{mailto:e.leonardi@lse.ac.uk}{e.leonardi@lse.ac.uk}. }
		
	}
	
	\vspace*{0.5cm}
	
    January 2025

	\vspace*{1cm}
	\begin{center}
		\begin{minipage}{1\linewidth}
			
			\textbf{Abstract} \hrulefill \vspace*{-0.0cm} \\
            How does the monetary and fiscal policy mix alter households' saving incentives? To answer these questions, we build a heterogenous agents New Keynesian model where three different types of agents can save in assets with different liquidity profiles to insure against idiosyncratic risk. Policy mixes affect saving incentives differently according to their effect on the liquidity premium- the return difference between less liquid assets and public debt. We derive an intuitive analytical expression linking the liquidity premium with consumption differentials amongst different types of agents. This underscores the presence of a transmission mechanism through which the interaction of monetary and fiscal policy shapes economic stability via its effect on the portfolio choice of private agents. We call it the \emph{self-insurance demand channel}, which moves the liquidity premium in the opposite direction to the standard \emph{policy-driven supply channel}. 
            Our analysis thus reveals the presence of two competing forces driving the liquidity premium. We show that the relative strength of the two is tightly linked to the policy mix in place and the type of business cycle shock hitting the economy. This implies that to stabilize the economy, monetary policy should consider the impact of the \emph{self-insurance} on the liquidity premium.

					\vspace*{-0.05cm}
		\noindent \hrulefill 
		\\
		\noindent \textsl{Keywords:}  monetary-fiscal interaction, liquidity, government debt, HANK
		\\
		\noindent \textsl{JEL Classification:} E12, E52, E62, E58, E63.

		\end{minipage}
		
	\end{center}
\end{center}
\vspace*{0.5cm}



\section{Introduction}
In recent years, the responses of fiscal and monetary policies to the financial crisis and the COVID-19 pandemic have led to a convergence of their operational boundaries. This has revived the attention on the long-term implications for economic stability and policy effectiveness of the fiscal-monetary policy mix and poses new challenges for the conduct of monetary policy in the presence of substantial fiscal expansions. 

At the same time, a burgeoning research agenda has highlighted the empirical differences in portfolios across the distribution of wealth and its relevance in terms of policy effectiveness and macroeconomic dynamics. In a seminal contribution, \cite{kaplan2018monetary} distinguish types of households by the prevalence of liquid or illiquid assets in their portfolios and describe the implications of different types of households for the transmission of monetary policy.
\cite{bayer2023liquidity} point to the importance of the liquidity channel of fiscal policy for its overall effect: the issuance of liquid government debt to finance discretionary government spending leads to a fall in the liquidity premium, defined as the return difference between less liquid assets and public debt. This, in turn, induces the households to save more in liquid assets that improve their ability to smooth consumption after negative shocks. We build on their work and analyze the transmission mechanism through which the interaction of monetary and fiscal policy shapes economic stability, i.e., via its effect on the portfolio choice of private agents. The purpose of this paper is to study how different combinations of active and passive policies affect the liquidity properties of the portfolios of different types of agents in the economy and what are the resulting challenges for the conduct of monetary policy.

For this, we use a New-Keynesian model with limited household heterogeneity in which three types of agents differ in their ability to trade in financial assets, in the spirit of \cite{Bilbiie:21}. The model features incomplete financial markets, on which agents can only trade liquid, nominal government bonds and an illiquid, real physical asset, i.e., capital. 
Our economy is made of \emph{capitalists}, who can trade in both markets; \emph{savers}, who can only adjust their liquid asset portfolio and cannot access the return from capital investments for consumption purposes; and \emph{hand-to-mouth} households, who cannot engage in the purchase of any asset, therefore relying on their labor income and previously accumulated government bonds for their consumption. Households are subject to idiosyncratic shocks that make them switch types according to an exogenous transition probability. When moving across types, households may only carry with them their government bonds. This characteristic defines the liquidity of this asset, contrary to capital.

Additionally, the presence of a fiscal authority that chooses the quantity of nominal tax revenues in a model with idiosyncratic uncertainty serves the further purpose of extending the range of policy mixes that we can consider in our analysis. \cite{hagedorn2018} shows that the demand for nominal bonds in incomplete-market economies combined with a fiscal authority that targets a nominal quantity leads to price-level determinacy independently of any interest rate rule. In our setup, the fiscal authority chooses the quantity of nominal taxes, and determinacy is attained by allowing nominal debt to adjust automatically to ensure that the government budget constraint is satisfied for any price level. The central bank then sets freely the nominal interest rate that clears the bond market, with no need to respond to any endogenous variable. Hence, the price level can be uniquely determined, jointly by monetary and fiscal policy, even when the interest rate is exogenously chosen.\footnote{Differently from here, \cite{hagedorn2018} and \cite{Bilbiie:21} have a set up where the fiscal authority chooses the quantity of nominal debt and fiscal revenues are then pinned down as a residual from the government budget constraint. From a price level determinacy perspective, this is equivalent to what we have here.} Hence, when fiscal policy is specified in nominal terms, shifts in the price level affect the real value of debt and thus affect real aggregate demand.\footnote{See online Appendix \ref{sec:Stability} for the stability analysis of our three-agents model and \cite{Bilbiie:21} and \cite{CD24} for detailed stability analysis of two-agent models with idiosyncratic risk, and nominal fiscal policy as in \cite{hagedorn2018} but without capital.} Crucially, this setup is different from the Fiscal Theory of the Price Level, which assumes that the government budget constraint must be satisfied by only one price level.

The first contribution of this paper is methodological. We extend the model of \cite{Bilbiie:21} to a three-agent (or states) setting, with different liquidity across two assets that are traded and held in positive net supply in equilibrium, so there is some notion of a “portfolio choice”, but still in a tractable way. This extension is needed to capture the ``self-insurance'' channel at the center of our analysis. This channel is present in quantitative HANK models like \cite{bayer2023liquidity} and \cite{auclert_intertemporal_2024}. We show that a two-agent set up is not enough to capture it and that our analytically tractable three-agents setup allows us to isolate this channel in the transmission of business cycle shocks, which is more difficult to do with a quantitative HANK model.

By considering different combinations of the monetary-fiscal regime together with the portfolio choice, we can indeed isolate this transmission mechanism that works via the liquidity premium. Due to the liquidity friction, government bonds are the preferred asset to build up a buffer stock of savings to partially insure against idiosyncratic uncertainty. Therefore, policy regimes that worsen the consumption ability of the hand-to-mouth by lowering their labor income or their bond income from previous states would increase the demand for self-insurance of the capitalist type, leading to a shift in their portfolios towards more liquid assets, \textit{ceteris paribus.} This is what we will label the ``self-insurance'' channel, and it works through asset demand.
In the analysis, we highlight the importance of this ``self-insurance'' channel for the dynamics of the liquidity premium, taken as encompassing the trade-off between investing in liquid and illiquid assets, deriving an equation that relates it directly to consumption differentials across types. To the best of our knowledge, we are the first to derive such an intuitive relation linking the liquidity premium to household consumption differentials in a HANK model.

Secondly, the policy regime will also determine the change in the supply of nominal government debt due to the change in the interest payment on its stock and the strength through which the government will curb the movement in debt. This is the standard ``supply'' channel, and it will have a further effect on the liquidity premium as the asset returns will have to adjust for markets to clear. This second channel would be present even in a representative agent/complete markets version of the model.
As a result, in our model, the liquidity premium is driven by the combination of two competing channels. One that works through asset demand (``self-insurance'') and the other via asset supply.
Different policy mixes will affect the relative strength of these two channels and, therefore, provide different results in terms of the relationship between the liquidity premium and the portfolio choice of households.

We then use the model to highlight the challenges for the conduct of monetary policy and answer questions regarding the aggregate implications of the combination of these two portfolio channels under different monetary and fiscal regimes. In our first experiment, we simulate and explore the implications of technology shock. Our objective is two-fold. On the one hand, we want to gauge the relative strength of the two channels highlighted above under different policy regimes. On the other hand, we want to explore the relative strength of the trade-off between the ``self-insurance'' channel via government bonds and the alternative ``insurance-through-investment'' channel. The latter comes from the fact that whilst a decline in liquid assets has a negative effect in principle because of the lower ability of agents to smooth consumption, a concomitant increase in capital investment may represent an indirect, general equilibrium form of insurance by potentially increasing the demand for labor and therefore the income of the hand-to-mouth type. 

Through this experiment, we show first that in regimes in which there is a large change in government debt, as when the monetary policy changes the interest rate strongly, and the fiscal policy does not intervene through taxes (active monetary, less passive fiscal regime), the supply channel reinforces the gap in the liquidity premium, inducing the profit motive of the capitalist to prevail. In this case, an increase in the liquidity premium leads capitalists to move towards less liquid assets, spurring an increase in investment. Regimes for which the change in government debt is more tamed, however, see a dominance of the self-insurance mechanism, whereby a worsening of the risky state induces a shift of capitalists towards the more liquid asset. Then, we show that in our model, in which we adopt a standard Cobb-Douglas production function, the ``investment-through-insurance'' channel discussed above is very weak due to the poor complementarity between the two productive assets. This implies that the building of more capital stock does not represent a viable substitute for liquid savings for agents in the hand-to-mouth state.

In our second experiment, we look at the effects of a fiscal stimulus with different combinations of the monetary/fiscal policy mix. An increase in government spending that produces a strong income effect for the hand-to-mouth agents reduces the ``self-insurance'' channel and induces capitalists to swap bonds for capital. At the same time, the fiscal stimulus increases the bond supply. We show that, once again, the relative strength of these two effects depends on the policy mix and has important implications for the conduct of monetary policy, especially in scenarios where fiscal policy is either too rigid or overly accommodative.  

When monetary policy is active, the supply channel dominates independently from the actions put in place by the fiscal authority. Whereas a less passive fiscal policy generates a larger cumulative fiscal multiplier, capitalists keep investing in bonds at the expense of capital. However, with passive monetary policy, a different picture emerges. The fiscal stimulus now induces a larger income effect on the hand-to-mouth, which pushes the liquidity premium further down and makes the ``self-insurance'' channel stronger. Under both fiscal policy scenarios, capitalists now substitute bonds with capital, which generates substantially larger fiscal multipliers compared to the active monetary policy simulations.

Finally, we look at the effects of a monetary policy shock. A monetary easing that reduces the interest rate has a direct effect on the liquidity premium, which falls. This induces a shift of the capitalists towards capital, which increases the investment in the economy. The increase in the price level reduces the real value of government debt, which in turn decreases the supply of bonds, making them even less attractive than capital. The strength of this supply channel, however, depends on the fiscal policy stance.

Comparing the simulations, under all experiments, of the proposed three-agent economy with the ones implied by a standard two-agent one demonstrates the importance of the ``self-insurance'' transmission channel we focus our analysis on. In the two-agent economy, the ``self-insurance'' channel is not active, and the liquidity premium between illiquid capital and liquid government bonds is only driven by the supply of public debt. As a result, in the two-agent economy, aggregate dynamics are mainly driven by monetary policy and do not depend on the fiscal policy stance. We show that this is not the case in the three-agent model, where aggregate dynamics now depend on the joint fiscal-monetary mix, given the presence of the ``self-insurance'' channel in affecting the dynamics of the liquidity premium.

Therefore, our analysis highlights that, in the current policy environment characterized by substantial fiscal expansions in response to the financial crisis and the COVID-19 pandemic, central banks should also consider the impact of the ``self-insurance'' channel on the liquidity premium when setting the monetary policy stance.

\subsection{Literature review}
Understanding monetary-fiscal policy dynamics is crucial to the formulation and implementation of effective policy measures aimed at promoting economic growth and stability. We see our paper contributing and merging two streams of the existing literature, i.e., the study of monetary and fiscal interactions and the aggregate consequences of households' portfolio choices, both of which are extremely prolific. 

First, we delve into the evolving discourse surrounding the interplay between monetary and fiscal policies. Initially, \cite{sargent1981} introduced the notion of "unpleasant monetarist arithmetic," illustrating the quandary faced by a central bank dedicated to curbing inflation while needing to accommodate inherently inflationary fiscal policies. Building upon this, \cite{sargent1984} extended the discussion to show how shifts in fiscal policy could undermine a central bank's commitment to maintaining low inflation. This research underscored the significance of fiscal policy expectations in influencing the efficacy of monetary policy, adding a crucial dimension to the discourse on the interaction of fiscal and monetary policies. Subsequently, \cite{leeper1991} introduced the concepts of "active" and "passive" monetary and fiscal policies and illustrated how both monetary and fiscal policies are endogenously determined within a model. Leeper's analysis highlighted the pivotal role of the chosen policy regime, whether fiscal or monetary policy is active or passive, in shaping the economy's response to shocks.

Building upon these foundations, a further series of studies explored the implications of policy rules and regime switching in the context of monetary and fiscal policy interactions \citep{davig2005, davig2011,bianchi2019,bianchi2020}. 

Instead of going down the route of regime-switching models, our contribution to this literature, leveraging the work of \cite{hagedorn2018}, consists of extending the static exploration of policy interactions beyond the traditional set of parameters' space. As discussed above, allowing the government to set fiscal policy in nominal terms induces price level determinacy in the model, even with passive monetary policy.

Then, our study adds to the dialogue on the aggregate consequences of households’ portfolio decisions in the presence of assets varying by their liquidity attributes. The role of government debt as liquidity, previously addressed by \cite{woodford1990public}, has recently gained renewed attention, as evidenced by studies such as \cite{bayer2023liquidity}, \cite{bilbiie2019capital} and \cite{Bilbiie:21}. Importantly, the ``self-insurance'' we study here is also present in \cite{bayer2023liquidity} and subsequent work like \cite{auclert_intertemporal_2024}.\footnote{More specifically, \cite{bayer2023liquidity} and \cite{auclert_intertemporal_2024} model capital illiquidity a la Calvo by assuming that participation in the capital market is random and, i.i.d. in the sense that only a fraction of households is selected to be able to adjust their capital holdings in a given period.}
While they strive to quantitatively identify the liquidity channel's influence on the effectiveness of fiscal policy using a model with a fully heterogeneous population of households, our study adopts an approach more closely aligned with that of \cite{bilbiie2019capital} and \cite{Bilbiie:21}, focusing on limited heterogeneity among household types. This method allows us to retain key elements of the larger, more complex models while clearly illustrating the ``self-insurance'' mechanism of transmission. 
In doing so, we also offer a methodological contribution by extending the \cite{Bilbiie:21}'s set-up to a three-agent (or state) setting. This gives us the possibility to derive an intuitive and analytical expression linking the liquidity premium to the consumption risk of agents and to show that a two-agent setup is not enough to capture the ``self-insurance'' channel we focus on in this analysis. Moreover, our analytical tractable setup allows us to isolate this channel in the transmission of business cycle shocks, which is more difficult to do with a quantitative HANK model.

The paper is structured as follows. In Section \ref{sec:model}, we outline the model. Then, in Section \ref{sec:experiments}, we explore the results of our experiments. In particular, we first look at a technology shock to answer the question of which of the channels outlined above prevails under different regimes, and then we move to the transmission of a fiscal and a monetary policy shock to study what the addition of the portfolio choice implies for the transmission of fiscal and monetary policies. We conclude by outlining how this analysis could be extended in future works.

\section{The benchmark model}\label{sec:model}
In this section, we present the model economy. As the main action takes place on the household side of the economy, we will be mainly focusing on detailing this. The production side follows the standard New-Keynesian specification \cite[see][]{gali2015monetary}, with CES final good producers and monopolistically competitive Rotemberg-pricing intermediate good firms. The government side will be modeled as a fiscal authority that chooses the quantity of nominal government debt and tax revenues, as in \cite{hagedorn2018}, to finance exogenous government spending.

\subsection{Households}
The household’s side is modeled in a way that can be defined as \cite{luetticke2018transmission} meets \cite{bilbiie2019capital}. This means that we are going to borrow the infrequent capital trading friction from the former and introduce it into the latter model of limited heterogeneity. In particular, we are going to focus on a three-agent model, in which households will switch between three states with exogenous transition probabilities governed by the matrix $\Lambda$ with generic component $\lambda_{i,j}$ for $(i,j)\in\{H,S,K\}^{2}$ as the transition probability of moving from state $i$ to state $j$. The difference among the three agents is going to be in their ability to access financial markets to insure against future income shocks. In particular, capitalists (indexed by $K$) can access capital markets in a way that allows them to adjust both their bond and capital holdings, whereas savers (indexed by $S$) will only be able to adjust and ripe the returns from government bonds. Furthermore, we assume that only capitalists are the firm's shareholders and receive their profits.
Finally, hand-to-mouth agents ($H$) will consume their labor income every period, as well as their accumulated savings income from bonds before transitioning to the hand-to-mouth state. It is in this sense that we define government bonds as a liquid asset and capital as an illiquid one, i.e., in terms of their consumption-smoothing insurance value to households.

We think of each type of agent as living on an island populated by their own type. Bonds can be carried across such islands, though they can only be adjusted on islands $K$ and $S$, and, as such, forward-looking agents will consider the consumption risk moving forward in their portfolio decisions. The benefits from holding capital, instead, can only be enjoyed on the $K$ island, therefore presenting a trade-off between the higher return commanded by its illiquidity and the desire to smooth consumption across states.\footnote{Online Appendix \ref{sec:K_nodrop} shows that results are robust to the case when we relax this assumption and allow capitalists leaving their island not to lose their accumulated capital wealth.}  
Given our tractable set-up, we will be able to highlight analytically the link between consumption risk and the liquidity premium in this economy.

\subsubsection{Population and financial accounting} 
We can think about the three types of consumers as inhabiting three distinct islands. We normalize the total population in the economy to 1 and denote with $\Pi_{i,t}$ for $i\in\{H,S,K\}$ the share of the population on each of the islands. Given our normalization, we have that $\Pi_{H,t}=1-\Pi_{S,t}-\Pi_{K,t}$. The evolution of each of these two shares follows the following laws of motion:
\begin{align}
\Pi_{K,t+1} &= \lambda_{K,K}\Pi_{K,t} + \lambda_{S,K}\Pi_{S,t} + \lambda_{H,K} (1-\Pi_{K,t}-\Pi_{S,t}) \label{lom_k}\\
\Pi_{S,t+1} &= \lambda_{K,S}\Pi_{K,t} + \lambda_{S,S}\Pi_{S,t} + \lambda_{H,S} (1-\Pi_{K,t}-\Pi_{S,t}) \label{lom_n}.
\end{align}
We look for the stationary distribution by setting $\Pi_{i,t+1}=\Pi_{i,t}=\Pi_{i}$ in the system above, which can be solved for the stationary shares as a function of the exogenous transition probabilities. From now onward, when referring to population shares, we mean the stationary ones, therefore omitting time subscripts. 

We follow the notation in \cite{Bilbiie:21} and \cite{bilbiie2019capital} and call $B_{t+1}^{j}$, for $j\in\{K,S\}$, the per-capita \emph{nominal} beginning-of-period $t+1$ bond holdings: after the consumption-saving choice, and also after changing state and pooling with the other agents of the same type. With a ``bold'' letter $\mathbb{B}$, we denote island-wide bond holdings at the beginning of the period. $Z_{t+1}^{j}$ instead are the end-of-period $t$ per-capita holdings of bonds of agent $j$, which the agents can choose after the consumption-saving choice but before they move across islands. Their respective evolution is as follows: 
\begin{align}
    \mathbb{B}^{K}_{t+1} &= \Pi_{K}B^{K}_{t+1} = \lambda_{K,K}\Pi_{K}Z^{K}_{t+1} + \lambda_{S,K}\Pi_{S} Z^{S}_{t+1} \label{lom_bk} \\
    \mathbb{B}^{S}_{t+1} &= \Pi_{S} B^{S}_{t+1} = \lambda_{K,S}\Pi_{K}Z^{K}_{t+1} + \lambda_{S,S}\Pi_{S} Z^{S}_{t+1} \label{lom_bn} \\
    \mathbb{B}^{H}_{t+1} &= (1-\Pi_{K}-\Pi_{S})B^{H}_{t+1} = \lambda_{K,H}\Pi_{K}Z^{K}_{t+1} + \lambda_{S,H}\Pi_{S} Z^{S}_{t+1} \label{lom_bu}.
\end{align}

\subsubsection{Household problem}
Each of the agent types will maximize the discounted sum of lifetime utility depending on the same specification as a function of the final consumption good and disutility from labor. Following the literature on the topic, we assume that there is a union that centralizes the wage-setting decision by pooling the labor supply of both types and allocates the hours equally across types, i.e. $N_t^{K}=N_t^{S}=N_{t}^{H}=N_t$. 

Therefore, agents will choose a path of consumption and, when possible, asset holdings to maximize the following period utility function:
\begin{equation}
\mathbf{E}_{0}\sum_{t=0}^{\infty} \beta^{t} \left(\frac{{C_{t}^{j}}^{1-\sigma}}{1-\sigma} - \psi\frac{{N_{t}^{j}}^{1+\varphi}}{1+\varphi}\right),
\end{equation}
for $j\in\{H,S,K\}$ subject to their respective flow of resources. For capitalists, that is
\begin{equation}
 P_t C^{K}_{t} + Z^{K}_{t+1} + P_{t}I^K_t = P_{t}W_{t}N^{K}_{t} + (1 + R^{b}_{t-1}) \frac{\mathbb{
 B}^{K}_{t}}{\Pi_{K}} + R^{k}_{t}P_{t} K^S_{t} + P_t \frac{D_{t}}{\Pi_{K}} - T_t^{K}- \tau^{K}, \label{bc_k}
\end{equation}
and similar for the purely bond savers (``savers'') is
\begin{equation}
P_t C^{S}_{t} + Z^{S}_{t+1} = P_tW_{t}N^{S}_{t} + (1 + R^{b}_{t-1}) \frac{\mathbb{B}^{S}_{t}}{\Pi_{S}} - T_t^{S}- \tau^{S}.\label{bc_n}
\end{equation}
Finally, as hand-to-mouth agents will not be able to trade bonds, their budget constraint will define their consumption as follows
\begin{equation}
P_t C^{H}_{t}= P_tW_{t}N^{H}_{t} + (1+R^{b}_{t-1}) \frac{\mathbb{B}^{H}_{t}}{\Pi_{H}}- T_t^{H}- \tau^{H}.\label{bc_u}
\end{equation}

$C^j_t$ is nominal consumption, $N^j_t$ hours of work, $K^K_t$ is capitalists capital stock, $I^K_t$ is investment in capital,  $W_t$ nominal wages, $R^b_t$ the risk free net nominal interest rate on bonds, $R^K_t$ the gross real rental rate of capital,  $D_t$ are economy-wide firms profits, $T^j_t$ are lump sum taxes, $\tau^j$ are steady state transfers to equate agents consumption, $\sigma$ the inverse of the intertermporal elasticity of substitution, $\varphi$ the inverse of the Frish elasticity, $\psi$ the disutility weight of labor and $\delta$ is capital depreciation.

We assume equal redistribution of the total tax revenue needed by the government.
Note that although type-H agents cannot trade in bonds, there will be a certain stock of this asset on the island as agents can carry these with them across type switches. 

We also assume that capital investment is subject to a convex adjustment cost $\iota$ so that capital accumulation reads:
\begin{equation}
    K^K_{t}=I^K_{t}\, \left(1-{{\iota}}\, \left(\frac{I^K_{t}}{I^K_{t-1}}-1\right)^{2}\right)+K^K_{t-1}\, \left(1-{{\delta}}\right). \label{iac}
\end{equation}

Since one of our focuses is fiscal policy, we introduce this adjustment cost to obtain fiscal multipliers in line with what is found in the literature, e.g., \citep{CantoreFreund:2021,HMM19}, under the active monetary policy regime, thereby making our analysis more realistic.

\subsection{Firms}
Since our model enriches the household side of the economy, we model firms according to a standard New Keynesian model with \cite{rotemberg1982sticky} adjustment costs. A continuum of monopolistically competitive ﬁrms indexed by $i \in [0,1]$ produce differentiated intermediate goods $Y_{t}(i)$ using labor $N_{t}(i)$ and capital $K_{t}(i)$ according to the following production function:
\begin{equation}
    F(A_t,N_t,K_t) =  (A_t  N_t(i))^{1-\alpha} K_t(i)^{\alpha} , \label{eq:prod}
\end{equation}
where $A_t$ is an AR(1) technology shock and $\alpha$ is the capital share of income. 
Firms seek to maximize their profits by optimally choosing their prices \(P_{t}(i)\) subject to the demand they face and price adjustment costs. The firm's problem is represented by the following dynamic optimization problem:

\begin{align*}
\max_{\{P_{t}(i)\}}& E_{0} \sum_{t=0}^{\infty} \Psi^K_{0,t} \left[ P_{t}(i) Y_{t}(i) - W_{t} N_{t}(i) - R^K_{t} K_{t}(i) - AC_{t} \right]\\
\text{s.t.}\,\, Y_{t}(i) &= \left(\frac{P_{t}(i)}{P_{t}}\right)^{-\epsilon} Y_{t},\\
Y_t(i) &=  (A_t  N_t(i))^{1-\alpha} + K_t(i)^{\alpha}-F ,
\end{align*}

where $\Psi^K_{0,t}=\beta^{t}\left(\frac{C^K_{t}}{C^K_0}\right)^{-\sigma}$ is the marginal rate of intertemporal substitution of capitalists, \(\epsilon > 1\) is the elasticity of substitution between different varieties of goods and $F$ are fixed costs in production to ensure 0 profits in steady state. 
We specify Rotemberg adjustment costs \(AC_{t}\) according to the standard quadratic representation:

\begin{align*}
AC_{t} = \frac{\xi}{2} \left( \frac{P_{t}(i)}{P_{t-1}(i)} - 1 \right)^{2} Y_{t},
\end{align*}

where \(\xi\) is a positive parameter that dictates the cost of adjusting prices. 

By optimizing this symmetric problem, the firms set the price to equate the real marginal cost ($MC_t$) to a markup over price, subject to adjustment costs. The first-order conditions for this problem generate the New Keynesian Phillips curve, providing the connection between inflation and output:

\begin{equation}
    (1-\epsilon)+\epsilon MC_t-\Pi_t \xi (\Pi_t-1)+\beta \Psi^K_{t,t+1} \Pi_{t+1}\xi(\Pi_{t+1}-1)\frac{Y_{t+1}}{Y_t}=0. \label{eq:nkpc}
\end{equation}

\subsection{Aggregation}
Aggregation of consumption and labor supply between the three types of households gives:

\begin{equation}
    C_t = \Pi_K  C^{K}_t + \Pi_S C^{S}_t + \Pi_H C^{H}_t,
\end{equation}
\begin{equation}
    N_t =  \Pi_K N^K_t + \Pi_S N^S_t + \Pi_H N^H_t.
\end{equation}

Aggregate capital and investment are given by:

\begin{eqnarray}
    K_t = \Pi_K K^K_t,\\
    I_t = \Pi_K I^K_t.
\end{eqnarray}

Total nominal government debt is:
\begin{equation}
    B_t=\Pi_K Z^K_t+\Pi_S Z^S_t.
\end{equation}

Finally the weighted sum of the three budget constraints \eqref{bc_k}, \eqref{bc_n} and \eqref{bc_u} gives the aggregate resource constraint.

\subsection{Policy block}

The Central Bank operates according to a standard Taylor rule of the form:
\begin{equation}
    \frac{1 + R^b_{t}}{1+R_b^{*}} = \left(\frac{1 + R^b_{t-1}}{1+R_b^{*}}\right)^{\rho_{R^b}}\left(\frac{\Pi_{t}}{\Pi^{*}}\right)^{(1-\rho_{R^b})\phi_{\pi}}M_t,
\end{equation}
where $M_t$ is an AR(1) monetary policy shock:
\begin{equation}
            \frac{M_t}{M^{*}}=\left(\frac{M_{t-1}}{M^{*}}\right)^{\rho_m}\epsilon_{m,t}.
\end{equation}
Variables with a $^{*}$ denote steady-state values.

The government is in charge of fiscal policy. In particular, it responds to a shock to nominal government expenditures, which follows a standard AR(1) process:
\begin{equation}
                \frac{G_t}{G^{*}}=\left(\frac{G_{t-1}}{G^{*}}\right)^{\rho_g}\epsilon_{g,t},
\end{equation}
by raising nominal lump sum taxes following a fiscal rule of the form
\begin{equation}
    \frac{T_t}{T^{*}}=\left(\frac{T_{t-1}}{T^{*}} \right)^{\rho_T } \left(\frac{B_{t-1}}{B^{*}} \right)^{(1-\rho_T)\gamma_T}\left(\frac{G_{t}}{G^{*}} \right)^{(1-\rho_T)\gamma_{TG}}\epsilon_{t,t}.
\end{equation}
$\epsilon_{t,t}$ represents a lump sum temporary tax/transfer shock.
Government nominal debt is therefore a residual pinned down by the inter-temporal nominal government budget constraint:
\begin{equation}
    G_t + B_{t+1} = (1+R_{t-1}^B)B_{t} + T_t.
\end{equation}

\subsection{Calibration}
Table \ref{tab:calibr} summarizes the calibration of the model. The value of the discount factor and capital depreciation are standard in quarterly models. We assume log-utility ($\sigma=1$) to highlight the precautionary savings coming purely from the liquidity channel and we set $\varphi=1$. Investment adjustment costs are set to 2.5. Hours in steady state assume that each agent works 1/3 of their time. We calibrate the steady-state share of government spending in output (20\%) and the annual debt-to-output ratio (57\%) to match the average for the US economy from 1984 to 2018. Rotemberg price adjustment costs are calibrated to match a frequency of price adjustment of 3.5 quarters. For cleaner intuition of the impulse responses, we assume no smoothing in the interest rate and taxes. In all simulations, tax response to government spending will remain fixed to $\gamma_{GT}=0.1$ as in \cite{GLV07}. The rest of the parameters of the monetary and fiscal policy rules will vary across exercises.  Passive fiscal policy will imply $\gamma_T=1$ while the less passive regime will have $\gamma=0.5$, which is the lower variable that ensures stability under all the scenarios analyzed below. Active monetary policy uses $\psi_{\pi}=1.2$ while passive $\psi_{\pi}=0.8$. Reducing further the response to inflation will generate much larger effects of fiscal policy.

Transition probabilities are calibrated such that capitalists remain in the $K$ island with probability $\lambda_{K,K}=0.8$ or move the $H$ one with probability $\lambda_{K,H}=0.02$. Hand-to-mouths instead have $\lambda_{H,K}=0.0541$ probability to become capitalists and savers $\lambda_{S,S}=0.95$ to stay savers. The rest of the transition probabilities are set accordingly to ensure that the population shares remain constant. This calibration follows \cite{BPT23} and \cite{Bilbiie:21} by implying that non hand-to-mouth agents have in beween 2\% ($\lambda_{K,H}=0.02$) and 4\% probability of becoming hand-to-mouth ($\lambda_{S,H}=0.0369$).\footnote{The calibration of the population weights together with the four transition probabilities mentioned above implies, assuming stationarity, the rest of the transition probabilities: $\lambda_{S,K}=0.0131$, $\lambda_{K,S}=0.18$, $\lambda_{H,S}=0.085$, and $\lambda_{H,H}=0.8609$. In online Appendix \ref{sec:Robustness}, we show robustness results to different calibrations of the transition probabilities.}
In line with data from the Survey of Consumer Finance, we calibrate to 20\% of the population living hand-to-mouth, 10\% capitalists while the rest is made of savers.\footnote{These population weights are also in line with calibrations in \cite{CantoreFreund:2021}, \cite{BPT23} and \cite{ORW23}.}

\noindent \textbf{Stability Analysis} As discussed in the introduction, \cite{hagedorn2018} set up with nominal fiscal policy permits us to extend the static exploration of policy interactions beyond the traditional set of parameters’ space. On the other hand, from \cite{Bilbiie:08}, we know that the presence of hand-to-mouth households affects the area of determinacy. Therefore, in online Appendix \ref{sec:Stability}, we perform a stability analysis of the model under the case of real and nominal fiscal policy. To be able to isolate as well the implications of the introduction of a third type of agent, we also perform the same stability analysis of a two-agent version of the model by setting the share of savers in the economy equal to 0 (and, therefore, $\Pi_K=0.8$). The results show that: i) the inverted Taylor principle (\emph{inverted aggregate demand logic} - \cite{Bilbiie:08}) is not present either in the two agents model with nominal fiscal policy or in the three agents model with real fiscal policy; ii) with real fiscal policy both models present four regions as in \cite{leeper1991} while with nominal fiscal policy we end up with just two regions. As fiscal policy becomes less and less passive ($\gamma_T$ approaching 0), we get instability, but otherwise, determinacy is attained independently of the monetary policy stance. Therefore, as in \cite{hagedorn2018}, allowing the government to set fiscal policy in nominal terms induces price level determinacy in the model even with passive monetary policy.


\begin{table}[htbp]
    \centering
    \begin{tabular}{lcl}
        \hline
        \textbf{Parameter} & \textbf{Value} & \textbf{Description} \\
        \hline
        $\beta$      & 0.99 & Discount factor \\
        $\sigma$     & 1    & Intertemporal elasticity of substitution \\
        $\delta$     & 0.025 & Capital depreciation rate \\
        $\varphi$    & 1  & Inverse of Frisch elasticity \\
        $\iota$      & 2.5  & Investment adjustment cost parameter \\
        $\Pi_K$      & 0.1  & Share of K-type households \\
        $\Pi_H$      & 0.2 & Share of hand-to-mouth households \\
        $\Pi_S$      & $1-\Pi_K-\Pi_H$ = 0.7 & Share of savers households \\
        $\lambda_{K,K}$ & 0.8   & Probability of a K-type staying K \\
        $\lambda_{K,H}$ & 0.02  & Probability of a K-type moving to H-type \\
        $\lambda_{K,S}$ & $1-\lambda_{K,K}-\lambda_{K,H}$ = 0.18  & Probability of a K-type moving to S-type \\
        $\lambda_{H,K}$ & 0.0541 & Probability of an H-type moving to K-type \\
        $\lambda_{S,S}$ & 0.95 & Probability of an S-type staying S \\
        $\lambda_{S,K}$ & $\frac{(1-\lambda_{K,K})\Pi_K- \lambda_{H,K}\Pi_H}{\Pi_N}$ = 0.0131 & Probability of an S-type moving to K-type \\
        $\lambda_{H,S}$ & $\frac{(1-\lambda_{S,S})\Pi_S- \lambda_{K,S}\Pi_K}{\Pi_H}$ = 0.0850 & Probability of an H-type moving to S-type \\
        $\lambda_{S,H}$ & $1-\lambda_{S,S}-\lambda_{S,K}$ = 0.0369 & Probability of an S-type moving to H-type \\
        $\lambda_{H,H}$ & $1-\lambda_{H,K}-\lambda_{H,S}$ = 0.8609 & Probability of an H-type staying H \\
        $\rho_m$       & 0.3 & Monetary policy shock persistence parameter \\
        $\rho_{R^b}$     & 0   & Interest-rate smoothing parameter \\
        $\phi_{\pi}$     & 0.8 or 1.2 & Taylor rule parameter \\
        $\rho_{g}$       & 0.9 & Government spending persistence parameter \\
        $\rho_{T}$       & 0   & Tax persistence parameter \\
        $\gamma_{T}$     & 0.5 or 1 & Tax response to debt \\
        $\gamma_{TG}$    & 0.1 & Tax response to government spending \\
        $\gamma_{G}$     & 0   & Government spending response to debt \\
        $\epsilon$         & 6   & Elasticity of substitution between goods varieties \\
        $\xi$         & 42.7 & Rotemberg price adjustment cost parameter \\
        $\rho_{a}$       & 0.75 & Technology shock persistence parameter \\
        $N^{*}$       & 0.33 & Steady-state labor supply \\
        $\Pi^{*}$       & 1    & Steady-state inflation rate \\
        $\frac{G^{*}}{Y^{*}}$ & 0.2 & Steady-state government spending-to-output ratio \\
        $\frac{B^{*}}{Y^{*}4}$ & 0.57 & Annualized steady-state debt-to-output ratio \\
        \hline
    \end{tabular}
    \caption{Calibration}
    \label{tab:calibr}
\end{table}

\subsection{The dynamics of the liquidity premium}

Before moving on to the model simulations in the following sections, we believe it is helpful to provide a characterization of the dynamics of the liquidity premium to understand better the determinants of the portfolio choice of the capitalists.
The liquidity premium in this economy is given by the expected return premium to hold the less liquid asset (capital) relative to the more liquid one (public debt).
Therefore we can write it as the ratio between the expected real gross return on capital and the expected real return on government bonds:
\begin{equation}
    LP_t=\mathbb{E}_t\frac{(R^K_{t+1}+(1-\delta)Q_{t+1})\frac{1}{Q_t}}{(1+R^B_{t})\frac{1}{\Pi_{t+1}}},
\end{equation}
where, due to the presence of investment adjustment costs, $Q_t$ is the price of capital goods relative to consumption goods.
For ease of exposition we call the real gross return of the two assets: $\mathcal{R}^K_{t+1}= \frac{R^K_{t+1}+(1-\delta)Q_{t+1}}{Q_t}$ and $\mathcal{R}^B_{t+1}=\frac{1+R^B_t}{\Pi_{t+1}}$. As a result the liquidity premium can be written as $LP_t=\mathbb{E}_t\frac{\mathcal{R}^K_{t+1}}{\mathcal{R}^B_{t+1}}$.

\begin{proposition}\label{prop_1}
The dynamics of the liquidity premium can be characterized up to the first order as follows: 
\begin{equation}
\hat{LP}_t=\beta \mathbb{E}_{t}\left[\Tilde{\mathcal{R}}^{K}_{t+1} - \Tilde{\mathcal{R}}^{B}_{t+1}\right] = -\sigma\mathbb{E}_{t}\left[\lambda_{K,S}(\hat{C}^S_{t+1}-\hat{C}^K_{t+1})+\lambda_{K,H}(\hat{C}^H_{t+1}-\hat{C}^K_{t+1})\right]. \label{liquidity_premium}
\end{equation}
\end{proposition}

In the above expression, ``$\Tilde{(\cdot)}$'' denotes that the variable has been linearized, as opposed to log-linearized (denoted by ``$\hat{(\cdot)}$''), in line with how the literature treats variables that are already expressed as percentages. The proof of this proposition, as well as further details on the analytics of the model, can be found in Appendix \ref{proofs}. 

Equation \eqref{liquidity_premium} helps us underpin in higher analytical details the mechanism that generates variations in the liquidity premium as related to self-insurance. The main mechanism is driven by the difference between the consumption of the capitalists and that of the other two types. The intuition behind this is that what determines the liquidity premium is the poor insurance quality of the capital asset. If the change in the consumption of the three agents is identical in every period, then the liquidity premium is neutralized, as there is no need for self-insurance to begin with. This will make the two assets perfect substitutes. In fact, in this case, the change in consumption will always be the same, regardless of which type the current capitalist will be in the next period. By contrast, if we consider a shock that induces a stronger response of the consumption of the non-hand-to-mouth type vis-à-vis those hand-to-mouth, we see that the last term in the equation above is likely to dominate. Assuming further that such deviation is positive, we see that it will generate a fall in the liquidity premium. Intuitively, such a shock makes the perspective of the risky hand-to-mouth state not as undesirable and, therefore, reduces the need for self-insurance. For this reason, the willingness to hold government bonds falls, and the real interest rate must increase for the market to clear. 
Through this mechanism, we see the importance of the interplay between monetary and fiscal policies in shaping the portfolio choice of agents with access to both capital and bond markets. Changes in the supply of government bonds, coupled with a change in their remuneration induced through monetary policy, will generate consumption inequality across the different categories, thereby affecting the need for insurance for the capitalist. This is what we will now study in the rest of the paper.

\section{The portfolio channel of monetary and fiscal policy interaction}\label{sec:experiments}

In this section, we report the dynamics of the model following a supply (\emph{technology}) and a demand (\emph{fiscal}) shock with a focus on different combinations of the monetary-fiscal policy mix. 

Following \cite{leeper1991}, for each of these shocks, we consider two scenarios for monetary policy: \emph{passive} ($\phi_{\pi}=0.8$) when the nominal interest rate responds less than one to one to inflation; \emph{active} ($\phi_{\pi}=1.2$) when the Taylor principle is satisfied. For each of these cases, we consider two scenarios for fiscal policy: \emph{passive} ($\gamma_T=1$) when nominal taxes respond one to one to the increase in nominal debt; \emph{less passive} ($\gamma_T=0.5$) when the reaction in taxes is more muted (i.e. fiscal policy decides to reduce as much as possible the increase in taxes to stimulate the economy).

\subsection{Technology shock}\label{sec:TS}
\begin{figure}[h!]
        \centering
        \includegraphics[width=\textwidth]{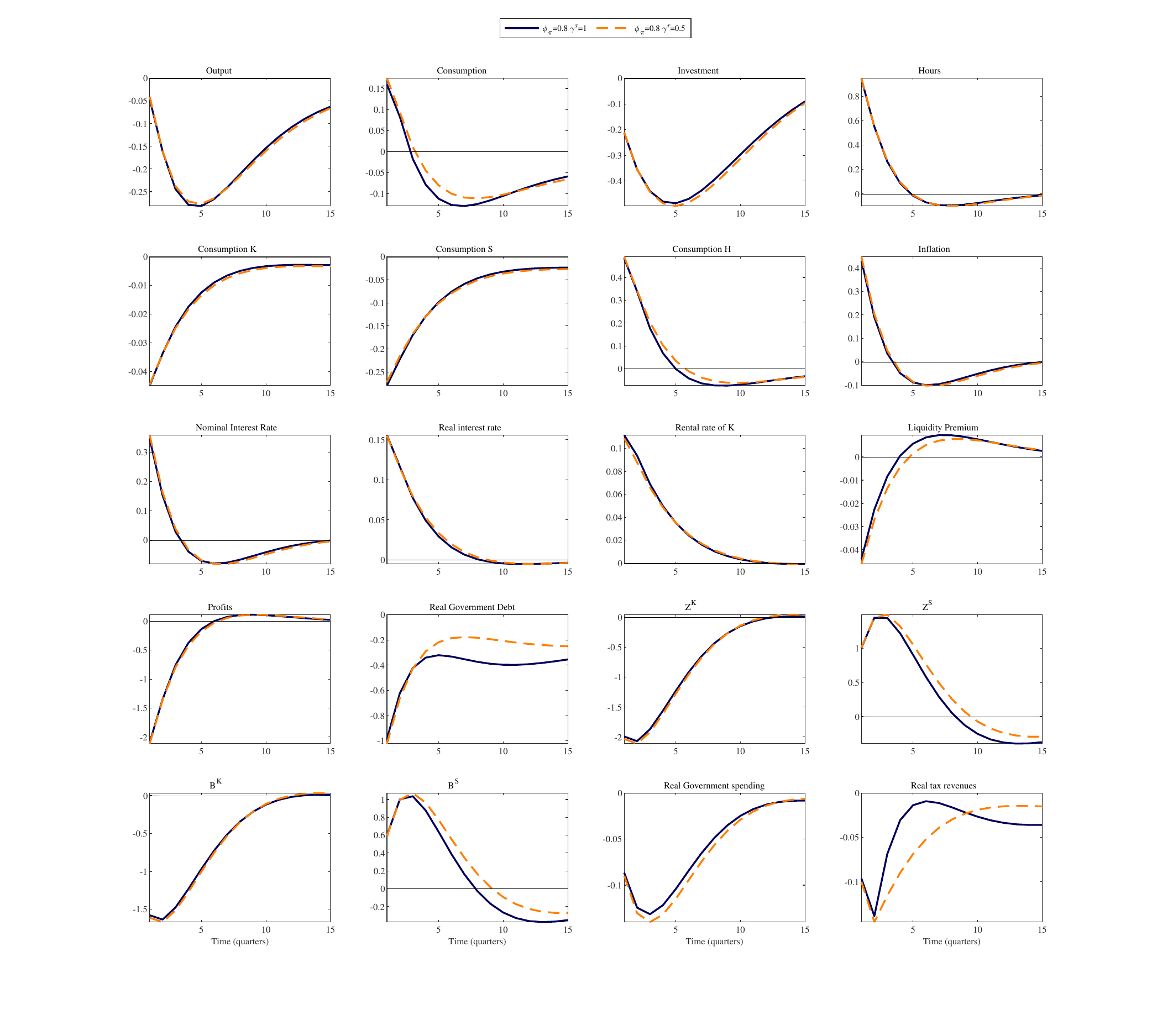}
        \caption{Impulse responses to a temporary 1\% decrease in A for the two fiscal policy regimes when monetary policy is passive ($\phi_{\pi}=0.8$).\\
        \begin{tiny}
            Note: All variables are expressed in real terms except for Hours, Inflation, and Nominal interest rate. All variables related to fiscal policy are in \% deviation from the steady state of output. The remaining variables are in \% deviations from their steady state. We plot the next period realized rental rate of capital ($R^K_{t+1}$). Consumption, Z's, and B's are island-wide figures (multiplied by the population sizes $\Pi$'s).
        \end{tiny}}
        \label{fig:3A_PM_epsZ}
    \end{figure}

\begin{figure}[h!]
        \centering
        \includegraphics[width=\textwidth]{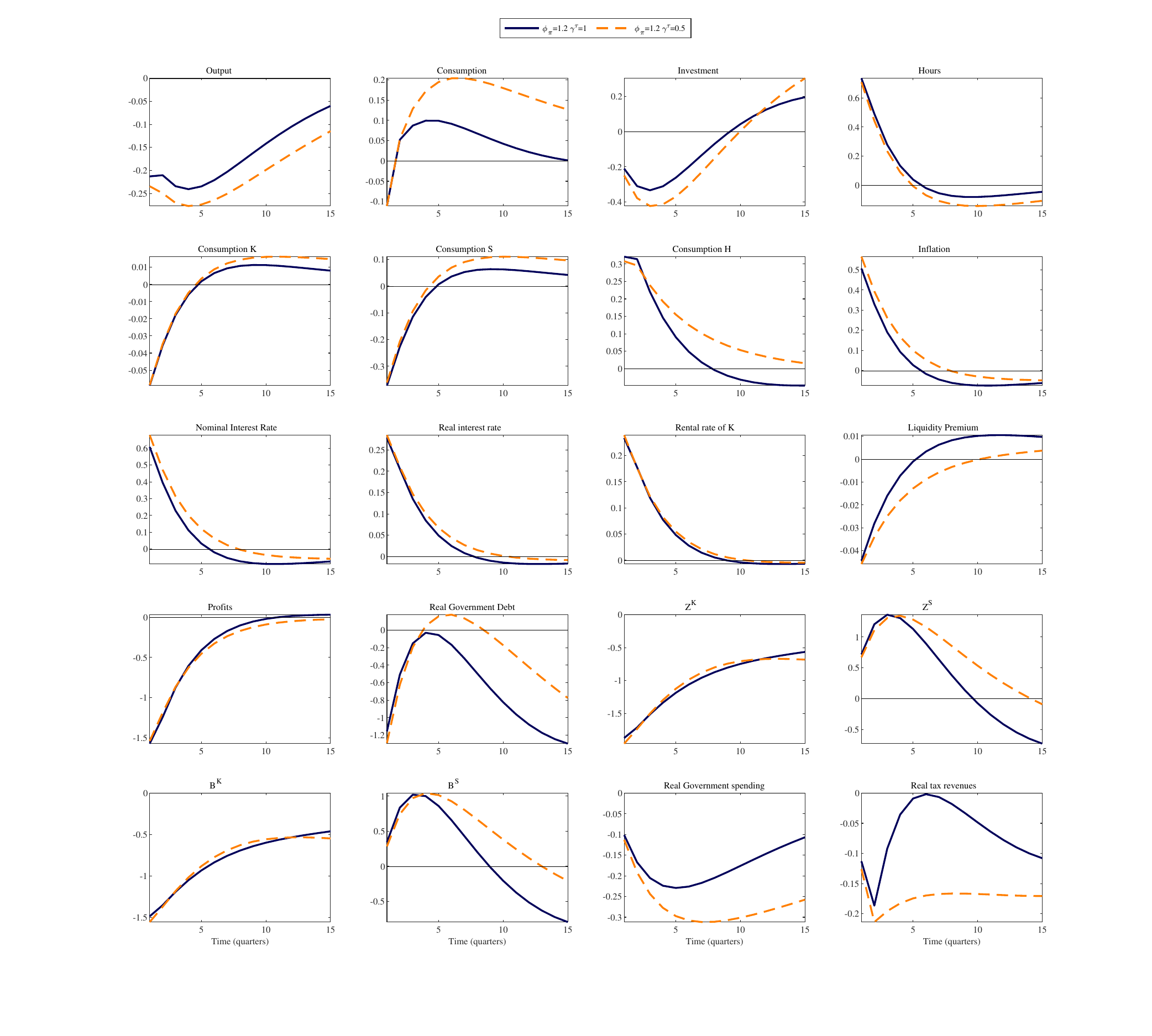}
        \caption{Impulse responses to a temporary 1\% decrease in A for the two fiscal policy regimes when monetary policy is active ($\phi_{\pi}=1.2$).\\
        \begin{tiny}
            Note: All variables are expressed in real terms except for Hours, Inflation, and Nominal interest rate. All variables related to fiscal policy are in \% deviation from the steady state of output. The remaining variables are in \% deviations from their steady state. We plot the next period realized rental rate of capital ($R^K_{t+1}$). Consumption, Z's, and B's are island-wide figures (multiplied by the population sizes $\Pi$'s).
        \end{tiny}}
        \label{fig:3A_AM_epsZ}
    \end{figure}

First, in Figure \ref{fig:3A_PM_epsZ} and Figure \ref{fig:3A_AM_epsZ}, we analyze the consequences of the portfolio channel on the transmission of a standard supply shock analyzed in the literature, i.e., a temporary and persistent $(\rho_a=0.75)$ technology shock $A_t$. In particular, each figure plots different fiscal regimes for the same monetary regime.

    The general transmission mechanism is the one established in the New Keynesian literature for this type of shock. A negative technology shock increases the real marginal costs, which determines an increase in inflation. Furthermore, it also generates a decrease in investments and a rise in labor hours. Importantly, we can see that a negative TFP shock relaxes the self-insurance mechanism we highlight in this paper. In particular, the rise in hours determines an increase in the consumption of hand-to-mouth, which decreases the desirability of liquid government bonds for self-insurance purposes (becoming hand-to-mouth becomes less undesirable). This can be seen from the portfolio choice of capitalists, who reduce their holdings of this type of asset. On the other hand, for savers who only have access to the bond markets, the standard interest rate channel is dominant, and they respond to the rise in interest rate by investing more and consuming less. Given the current calibration, we see that the bond holdings of hand-to-mouth follow a similar path as that of savers, which further improves the rise in consumption of the former. In line with the lower demand for self-insurance of the capitalists, the liquidity premium, which quantifies the insurance value of government bonds as pointed out in Proposition 1, decreases. 

    By moving to a comparative analysis across regimes, we can see how different combinations affect the portfolio channel and, therefore, the aggregate dynamics in the economy. In particular, a comparison across Figures \ref{fig:3A_PM_epsZ} and \ref{fig:3A_AM_epsZ} shows how a more active monetary policy in response to an inflationary technology shock induces stronger and more persistent movements in output, and consequently, a smaller increase in hours. Importantly, this logic is not peculiar to technology shocks but will hold for all supply shocks that move inflation and output in opposite directions in the context of a monetary policymaker that does pure inflation targeting, such as a cost-push shock: a stronger monetary policy response to tame inflation is going to have a destabilizing effect on output which will limit the response of hours.
    This translates, in general, into a lower increase in hand-to-mouth consumption and a higher decline in capitalist consumption. As a consequence, the decline in the liquidity premium, on impact, is similar to the one of Figure \ref{fig:3A_PM_epsZ}. However, its persistence is much higher now when less passive fiscal policy is at play.


This is because the interaction of monetary and fiscal policy brings a further mechanism at play, which was not evident in the case of passive monetary policy. After an initial decrease due to the shift in capitalists’ portfolios and counterbalanced by a decrease in taxes for government budget constraint to hold, real government debt experiences a rise for two reasons, i.e., the increasing demand of capitalists and the increased interest payments on the existing stock. The strength of this rise is driven by the monetary reaction as well as how strongly the government tries to stabilize it through taxes. In particular, if the fiscal policy does not react strongly to the change in debt, the stock of government debt is going to follow a more strongly increasing path, coupled with an active monetary policymaker that changes rates aggressively and raises the interest payments on government debt. This is evident in Figure \ref{fig:3A_AM_epsZ}. This strong increase in the supply of real government debt has a feedback effect on the liquidity premium, which takes longer to recover. 

Furthermore, we can see how different fiscal policy stances affect the liquidity premium and, consequently, capital investment and, ultimately, output. When monetary policy is active, a passive fiscal policy delivers a lower decline in investment and output.
This supply effect is largely non-present in Figure \ref{fig:3A_PM_epsZ}, where we see that with passive monetary policy, the fiscal stance does not affect the macroeconomic variables. 
Our simulations also show that, for a supply shock, the trade-off between self-insurance and ``insurance-through-investment''\footnote{As explained in the introduction, the ``insurance-through-investment'' mechanism comes from the fact that whilst a decline in liquid assets has a negative effect in principle because of the lower ability of agents to smooth consumption, a concomitant increase in capital investment may represent an indirect, general equilibrium, form of insurance, by potentially increasing the demand for labor and therefore the income of the hand-to-mouth type. } is resolved in favor of the former. In fact, following a negative TFP shock, a decrease in investment increases the demand for labor and creates a substitution effect that improves the position of the hand-to-mouth. This might be overturned in the presence of a more general production function, e.g., of the CES form, which would allow for complementarity between capital and labor.

Finally, we want to show the importance of the portfolio channel via the liquidity premium for the transmission of supply shocks, compared to the standard transmission mechanism. To do so, in online Appendix \ref{sec:2A}, we provide simulations of the two-agent version of the model (by setting $\Pi_S=0$, and $\lambda_{K,K}=0.98$ as in \cite{bilbiie2019capital} and \cite{BPT23}) where there are no savers and therefore there are no meaningful portfolio choices of capitalists affecting the demand for Government debt. Results show that in the two-agent economy, the dynamics are mainly driven by the monetary authority independently of the fiscal stance. This is because the self-insurance channel is not present in this economy, and therefore, the liquidity premium barely moves.

\subsection{Fiscal Shock}\label{sec:FS}
How does fiscal policy transmit with different combinations of the monetary/fiscal policy mix?
We answer this question by looking at the shock to government spending and focusing our attention on the effect on real public debt when different combinations of monetary and fiscal policy lead to different behaviors of inflation.

\begin{figure}[h!]
    \centering
    \includegraphics[width=\textwidth]{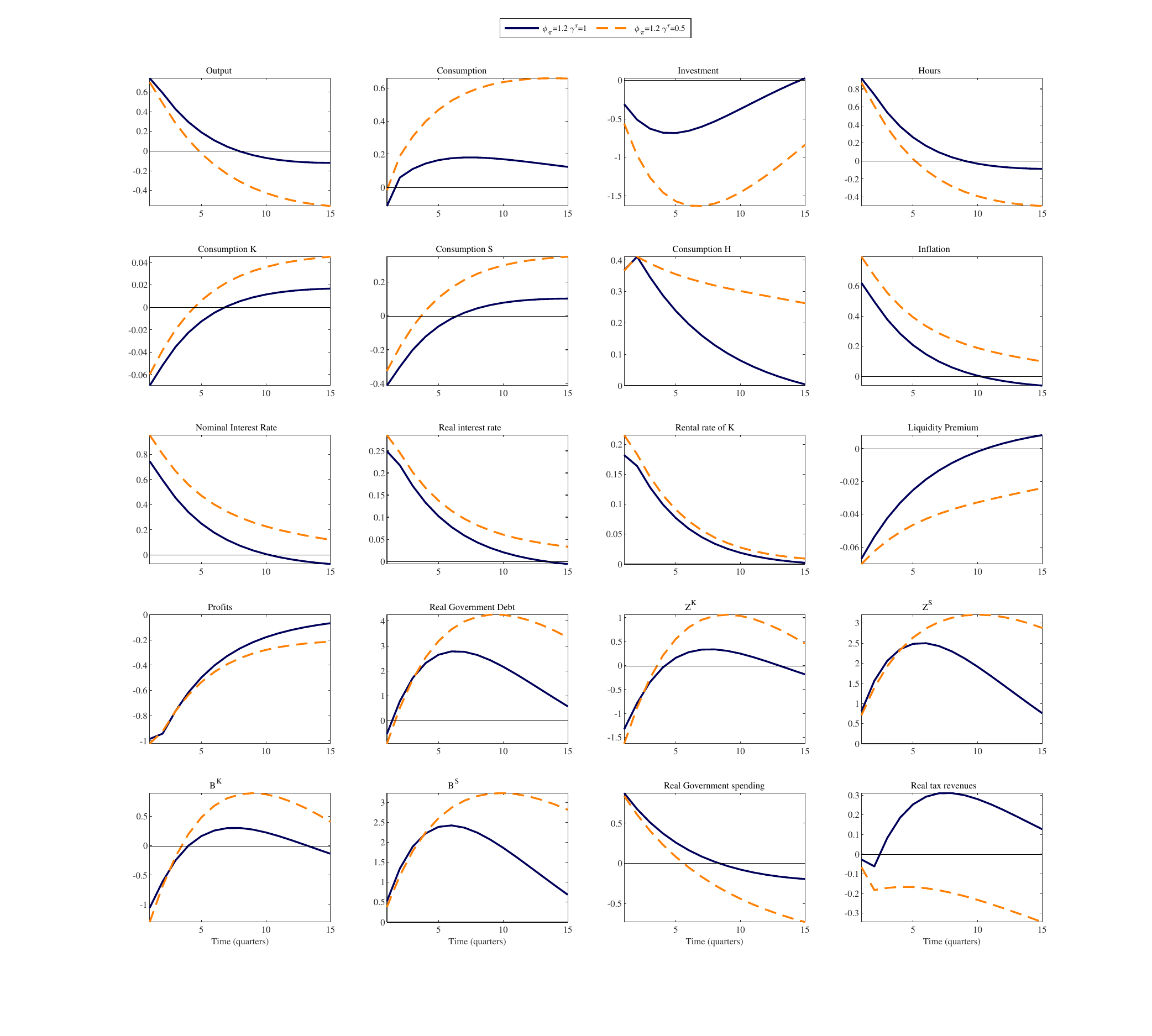}
    \caption{Impulse responses to a temporary 1\% increase in $G^N$ for the two fiscal policy regimes when monetary policy is active ($\phi_{\pi}=1.2$).\\
    \begin{tiny}
        Note: All variables are expressed in real terms except for Hours, Inflation, and Nominal interest rate. All variables related to fiscal policy are in \% deviation from the steady state of output. The remaining variables are in \% deviations from their steady state. We plot the next period realized rental rate of capital ($R^K_{t+1}$). Consumption, Z's, and B's are island-wide figures (multiplied by the population sizes $\Pi$'s).
    \end{tiny}}
    \label{fig:3A_AM}
    \end{figure}

    Figure \ref{fig:3A_AM} shows the responses of key variables to a persistent ($\rho_g=0.9$) increase in G of 1\% in terms of output for the case of active monetary policy.\footnote{In online Appendix \ref{sec:Tshock}, we also report simulations to a temporary and uniform lump-sum transfer shock.} 
    The increase in government purchases raises the level of aggregate demand in the economy. Following the standard New Keynesian narrative, firms operating under monopolistic competition raise their prices, however, given nominal rigidities, this change is insufficient to fully restore the original equilibrium. The labor demand curve shifts outwards, hours worked and output rise; this implies an increase in the real wage (not shown).  In contrast to the representative-agent paradigm, the presence of a high marginal propensity to consume agents ($H$) means that they will see a boost in their disposable income and raise their consumption. In contrast, capitalists and savers who have access to financial markets will reduce their consumption. As a result, aggregate consumption is slightly crowded out on impact, but it quickly turns positive, and its magnitude depends on the fiscal policy stance.
    Capital investment, instead, is persistently crowed out in line with the standard transmission mechanism in the presence of investment adjustment costs.

    The effect on real debt can be decomposed into demand and supply-side effects. From the demand side, the strong income effect that pushes up consumption of the hand-to-mouth makes the $H$ state less undesirable and, therefore, lowers the liquidity premium, making bond and capital closer substitutes. This induces a shift of the capitalists out of the bond market and into the capital market as their demand for \emph{insurance} via bonds is reduced ($Z^K$ declines for the first three or four quarters, and the initial drop in investment is contained). However, there is a contrasting supply-side effect on capital vs bond investment. This is due to the increase in nominal debt issuance generated by the increase in government spending and by the increase in real debt due to the jump in inflation and the real interest rate. This standard ``interest rate'' channel makes bonds more attractive with respect to capital.

    The relative strength of these two effects depends on the policy mix. When monetary policy is active, the supply side effect dominates (Figure \ref{fig:3A_AM}). When fiscal policy is passive (solid blue line), the rise in tax revenues corresponds to a smaller increase in inflation, which translates to a smaller decline and faster recovery in the liquidity premium compared with the less passive fiscal policy case (dashed orange line). If the self-insurance motive of capitalists dominates, we would observe a smaller drop in investment in the less passive fiscal policy case, while our simulations show the opposite. This is evidence that the supply-side effect is driving the response of real debt to the fiscal shock.
    
    The different behavior of inflation and the liquidity premium also explain the different fiscal multipliers under the two fiscal policy scenarios. On impact, the fiscal multiplier is slightly larger under passive fiscal policy, but the difference widens when looking at the cumulative effect.

    \begin{figure}[h!]
        \centering
        \includegraphics[width=\textwidth]{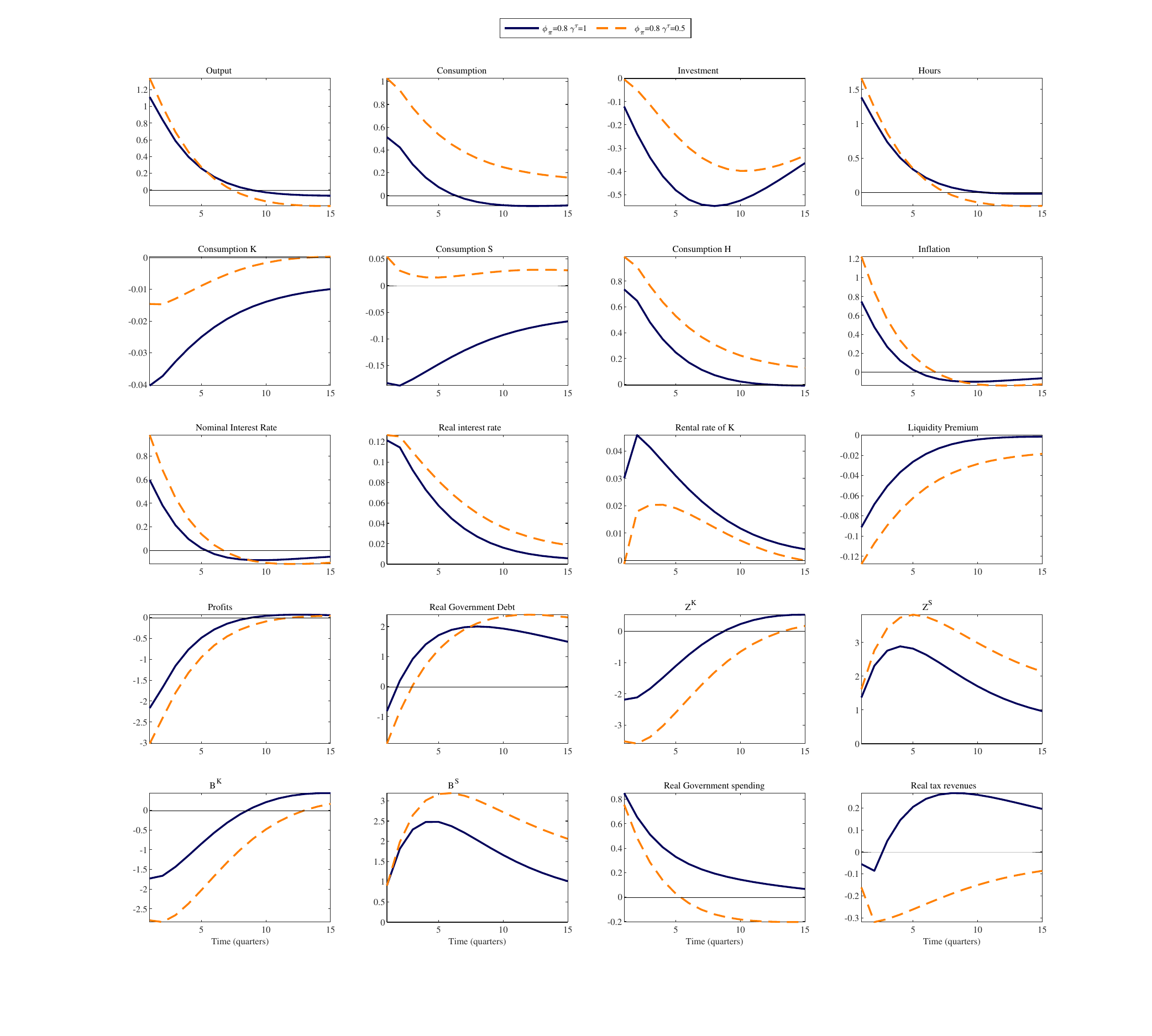}
        \caption{Impulse responses to a temporary 1\% increase in $G^N$ ffor the two fiscal policy regimes when monetary policy is passive ($\phi_{\pi}=0.8$).\\
          \begin{tiny}
            Note: All variables are expressed in real terms except for Hours, Inflation, and Nominal interest rate. All variables related to fiscal policy are in \% deviation from the steady state of output. The remaining variables are in \% deviations from their steady state. We plot the next period realized rental rate of capital ($R^K_{t+1}$). Consumption, Z's, and B's are island-wide figures (multiplied by the population sizes $\Pi$'s).
    \end{tiny}}
        \label{fig:3A_PM}
    \end{figure}
    
    Figure \ref{fig:3A_PM} shows the responses for the case of passive monetary policy. Compared to the case of active monetary, the impact of the shock is qualitatively similar for many variables, as the standard New-Keynesian narrative also applies in this case. Quantitatively, however, we observe a substantially larger fiscal multiplier, a larger increase in inflation, and a smaller rise in the real rate.  When the central bank lets inflation increase by a larger amount, the demand effect of the fiscal shock is magnified. This induces a stronger income effect on the hand-to-mouth, which reduces even further the self-insurance motive of capitalists. Their shift away from the bond market is much more pronounced and persistent. Therefore, in this case, the demand effect now dominates over the supply of new public debt. Under both fiscal policy scenarios, capitalists substitute bonds for capital ($Z^K$ is negative and persistent).
    When fiscal policy is less passive, we even observe a decline in real government debt for a few quarters.
    Finally, we also notice how, under the less passive fiscal policy, the increase in inflation is so large that it generates a substantial decline in real tax revenues, which frees up disposable income and generates an increase in consumption for both capitalists and savers.

    Another way to highlight the importance of the demand side effect coming from the reduced insurance motive of capitalists is to look, as we did for the technology shock case, at the same impulse responses in the two-agent version of the model where only the supply side effect is at play (see online Appendix \ref{sec:2A}).  Simulations show once again that, given the monetary policy stance, fiscal policy does not have a visible impact on macroeconomic variables in the two-agent economy.
    Let's consider the passive monetary policy scenario in Figure \ref{fig:2A_PM} for example. In this model, capital and bonds are perfect substitutes, and therefore, the capitalists do not substitute bonds for capital, as in the three-agent economy. This leads to a larger decline in investment in the two-agent model compared to the less passive fiscal policy case in the three-agent economy. Note also the standard decline in the consumption of capitalists, which further contributes to generating a smaller fiscal multiplier in the two-agent setup.
    
    Another way to compare different policy mixes within and across models is indeed to look at the fiscal multipliers. Table \ref{tabDSGE_Multipliers} reports the impact and cumulative multipliers (over 15 quarters) for the three-agent model and the two-agent version. The comparison between the two models highlights once more the importance of the portfolio channel in the transmission of fiscal shocks. In the absence of the demand side ``self-insurance'' channel, the supply side effect dominates, and the fiscal policy stance has no impact on the multipliers. Instead, in the three-agent model, the multipliers depend on the joint behavior of monetary and fiscal policy. In particular, the fiscal multiplier is larger when monetary policy is passive and fiscal policy is less passive. This is because the demand side effect of the fiscal shock is magnified by the reduced insurance motive of capitalists. 

    In sum, as we saw for the technology shock, the comparison between our model and the two-agent version highlights the importance of the portfolio channel in the transmission of fiscal shocks. In the absence of the demand side ``self-insurance'' channel, the supply side effect dominates, and the fiscal policy stance has a limited impact on the macroeconomic variables.

    \begin{table}
        \begin{center}
            \begin{spacing}{1.15}
                \begin{threeparttable}
                    \begin{tabular}{lcccccc}
                        \FL
                          \multicolumn{6}{c}{Three agents model}  \NN 
                        \FL
                        &&  \multicolumn{2}{c}{Passive Monetary} &  \multicolumn{2}{c}{Active Monetary}  \NN 
                        \FL
                        && $\gamma_T=1$ & $\gamma_T=0.5$  &$\gamma_T=1$ & $\gamma_T=0.5$   \NN \FL
                        Impact multiplier & &1.31 &	1.77 &	0.84 &	0.83 \NN 
                        Cumulative multiplier & & 0.54 &	0.57 &	0.35 &	0.00  
                        
                        \LL
                        \multicolumn{6}{c}{Two agents model}  \NN 
                        \FL
                        &&  \multicolumn{2}{c}{Passive Monetary} &  \multicolumn{2}{c}{Active Monetary}  \NN 
                        \FL
                        && $\gamma_T=1$ & $\gamma_T=0.5$  &$\gamma_T=1$ & $\gamma_T=0.5$   \NN \FL
                        Impact multiplier & &  1.14 &	1.14 &	0.89 &	0.89      \NN 
                        Cumulative multiplier & & 0.51 &	0.51 &	0.30 &	0.30     
                        
                        \LL
                    \end{tabular}
                \end{threeparttable}
            \end{spacing}
            \caption{Fiscal multipliers}
            \label{tabDSGE_Multipliers}
            \medskip
            \begin{minipage}{1\textwidth} 
                {\footnotesize \emph{Notes:} This table summarizes the output effects of a government spending shock according to different models: the first two rows refer to the three agents model described in section \ref{sec:model}, the bottom two rows refer to the two-agent version where we set $\Pi_K=0.8$ and $\Pi_S=0$.
                }
            \end{minipage}
        \end{center}
    \end{table}

\subsection{Monetary Shock}\label{sec:MP}
As a final exercise, we want to study how monetary policy transmission is affected by different combinations of the monetary/fiscal policy mix. Hence, we look at the effect of a monetary policy shock on the economy. 

\begin{figure}[h!]
\centering
\includegraphics[width=\textwidth]{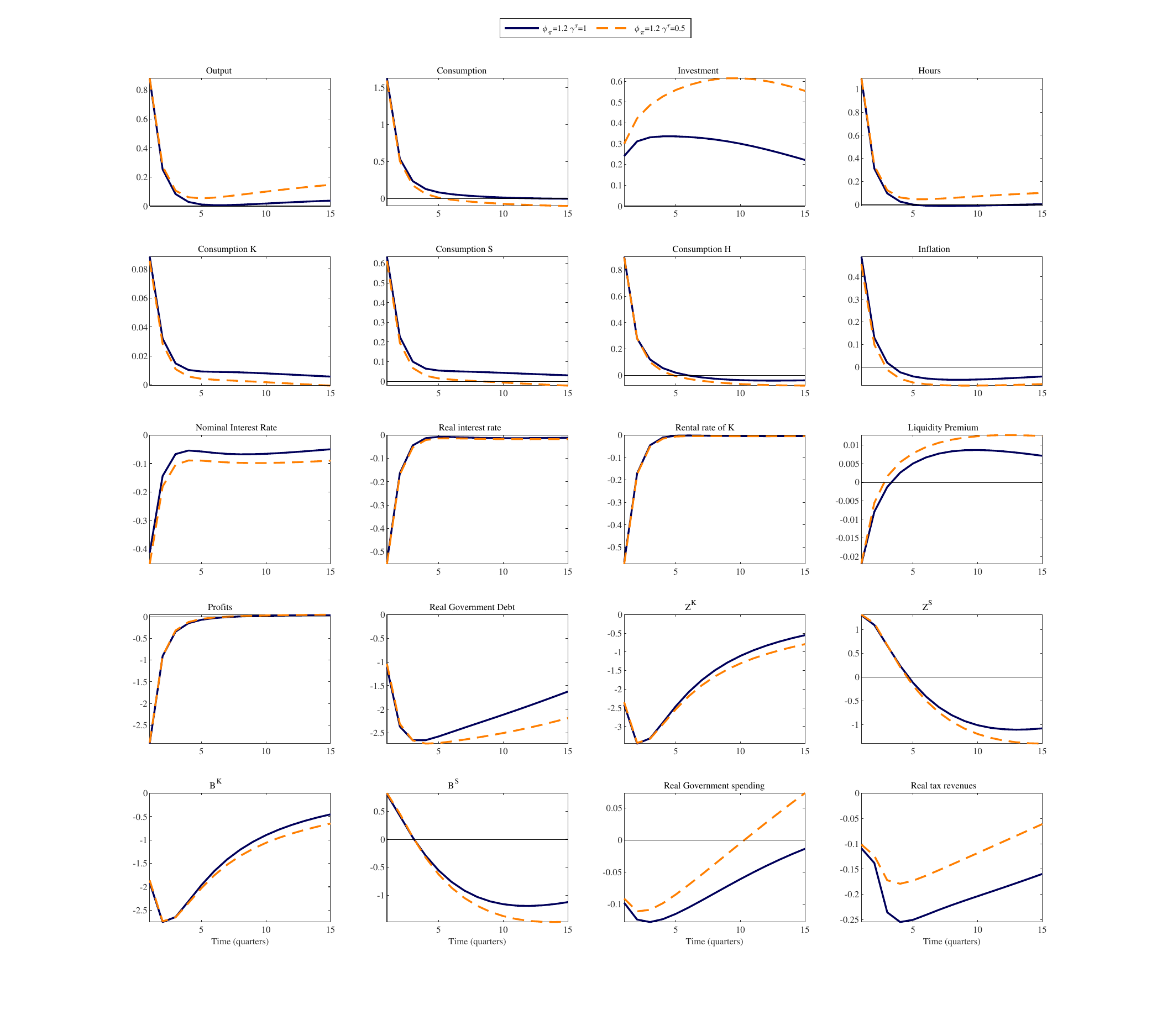}
\caption{Impulse responses to a temporary 1\% increase in M for the two fiscal policy regimes when monetary policy is active ($\phi_{\pi}=1.2$).\\
\begin{tiny}
    Note: All variables are expressed in real terms except for Hours, Inflation, and Nominal interest rate. All variables related to fiscal policy are in \% deviation from the steady state of output. The remaining variables are in \% deviations from their steady state. We plot the next period realized rental rate of capital ($R^K_{t+1}$). Consumption, Z's, and B's are island-wide figures (multiplied by the population sizes $\Pi$'s).
\end{tiny}}
\label{fig:3A_AM_epsM}
\end{figure}

Figure \ref{fig:3A_AM_epsM} shows the responses of key variables to a persistent ($\rho_m=0.3$) increase in $M$ for the case of active monetary policy.
The standard NK transmission of monetary policy is at play. The decrease in the nominal interest rate induces an increase in the marginal cost and inflation as well as consumption of all agents, hours worked, output, and investment. Importantly, we note how this shock decreases the self-insurance motive because the consumption of capitalists increases less than the consumption of hand-to-mouth on impact. As a result, this induces capitalists to substitute investment in bonds with investment in physical capital, therefore pushing aggregate investment further up. 
The increase in the price level has a negative impact on real debt and spending. This means that the standard supply channel makes bonds even further less attractive than capital. This second channel, as well as the effect on real tax revenues and obviously on inflation, depends on the fiscal stance.
With less passive fiscal policy, we observe a larger increase in investment, which translates into a larger and more persistent output above the trend.

\begin{figure}[h!]
    \centering
    \includegraphics[width=\textwidth]{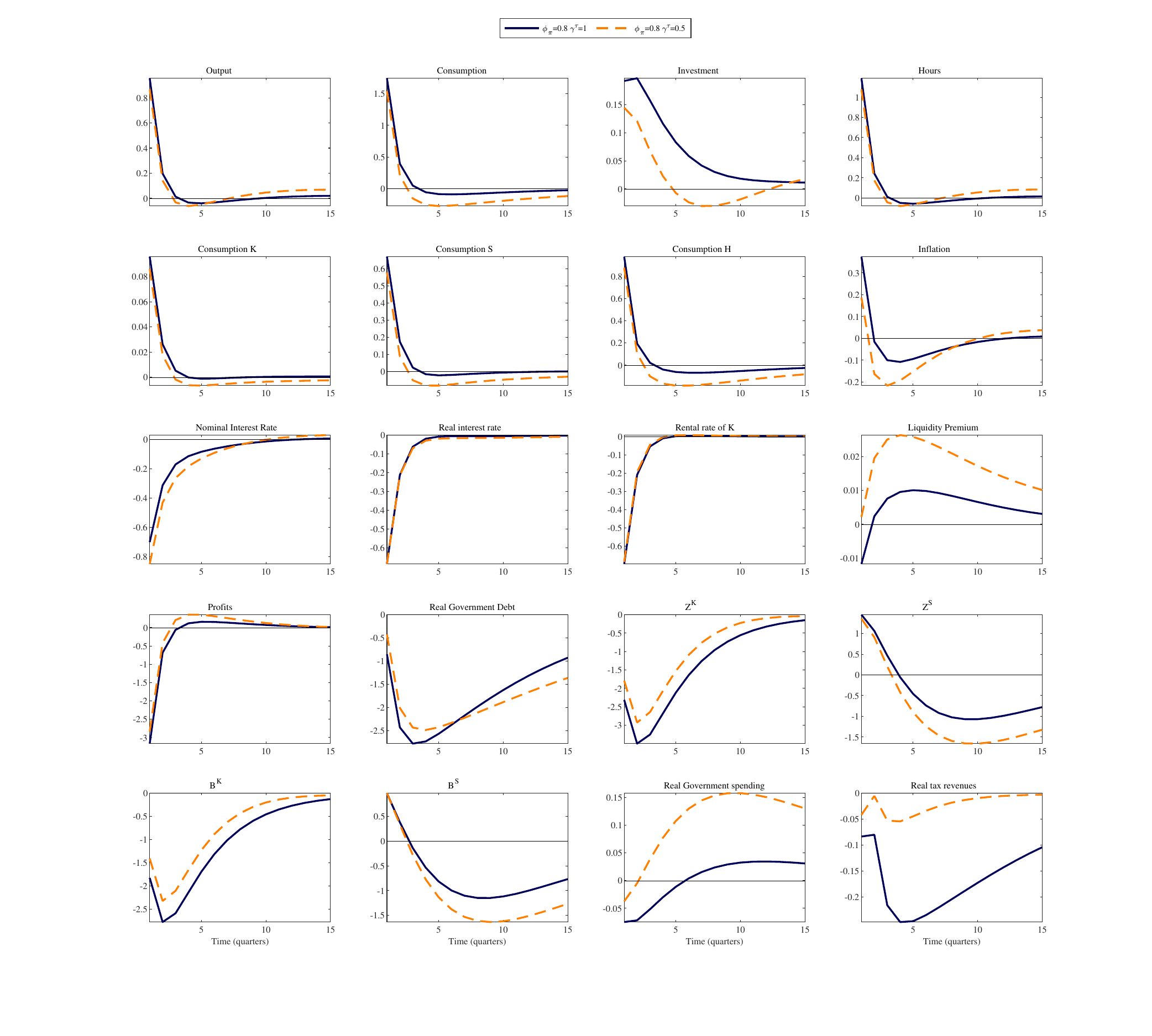}
    \caption{Impulse responses to a temporary 1\% increase in M for the two fiscal policy regimes when monetary policy is passive ($\phi_{\pi}=0.8$).\\
        \begin{tiny}
        Note: All variables are expressed in real terms except for Hours, Inflation, and Nominal interest rate. All variables related to fiscal policy are in \% deviation from the steady state of output. The remaining variables are in \% deviations from their steady state. We plot the next period realized rental rate of capital ($R^K_{t+1}$). Consumption, Z's, and B's are island-wide figures (multiplied by the population sizes $\Pi$'s).
\end{tiny}}
    \label{fig:3A_PM_epsM}
\end{figure}

Turning to the case of passive monetary policy (Figure \ref{fig:3A_PM_epsM}), we can see the standard NK transmission of monetary policy at play again. However, now the effect of the fiscal stance flips compared to the previous case. When monetary policy is passive, a less passive fiscal policy substantially reduces inflation after the first period. This, in turn, negatively affects the disposable income of hand-to-mouth via its effect on real taxes and pushes the liquidity premium up, resulting in aggregate investment and output being below the case with a more passive fiscal policy. 

As for the previous two shocks, we also show simulations for the two-agent version of the model in online Appendix \ref{sec:2A}. The results, once more, confirm that the fiscal policy stance does not affect the macroeconomic variables in the two-agent economy. This is because of the absence of the self-insurance motive of the capitalists.

\clearpage

\section{Conclusions}
This paper studied how the reciprocal interplay of fiscal and monetary policy affects the liquidity properties of the portfolios of different types of households. In doing so, we built upon the burgeoning research agenda, focusing on the empirical differences in portfolios across the wealth distribution and its impact on policy effectiveness and macroeconomic dynamics.

Utilizing a New-Keynesian model with limited household heterogeneity, we considered the transmission mechanism arising from the demand side of assets that we label the ``self-insurance'' channel. Through this channel, monetary and fiscal policy interact with the portfolio choices of private agents. This analysis revealed that the ``self-insurance'' channel operates opposite to the standard one coming from the ``supply'' of public debt, driving the liquidity premium.
 The presence of these two competing forces determines the behavior of the liquidity premium under different monetary and fiscal policy regimes. It shapes the relationship between this premium and the portfolio choices of households.

Our first experiment illustrated how these channels manifest and interact in response to a standard technology shock. We found that when there is a large change in government debt, the supply channel enhances the gap in the liquidity premium. 
On the contrary, in regimes where changes in government debt are more tempered, the ``self-insurance'' mechanism dominates. Moreover, our model suggested that the ``insurance-through-investment'' motive is weak due to the poor complementarity between the two productive assets, indicating that the building of more capital stock, via its general equilibrium effect on labor income, does not serve as a viable substitute for liquid savings for agents in the hand-to-mouth state.

In our second experiment, we explored the effects of a fiscal stimulus within different combinations of the monetary/fiscal policy mix. The relative strength of the self-insurance and supply channels was found to be dependent on the policy mix. Under an active monetary policy, the supply effect dominated, while under a passive monetary policy, a larger income effect on hand-to-mouth agents made the ``self-insurance'' channel stronger.

In our third and final experiment, we looked at the propagation of a monetary policy shock under different policy mixes. We found that the fiscal stance could affect the liquidity premium and, consequently, the investment and output dynamics. In the case of an active monetary policy, a reduction in the policy rate that stimulates the economy reduces the self-insurance motive of capitalists, leading to a substitution of bonds for capital. The resulting increase in the price level reduces real debt and spending, and, as a result, the supply channel makes bonds even less attractive than capital. This effect is magnified when fiscal policy is less passive. In the case of a passive monetary policy, instead, the effect of the fiscal stance on the propagation of monetary policy shocks flips.

We also showed how comparing the simulations of the three-agent economy with those implied by a standard two-agent one demonstrates the quantitative importance of the transmission channel studied in this paper. In a two-agent economy, the 
absence of the ``self-insurance'' channel makes monetary policy the main driver of aggregate dynamics, independently of the fiscal policy stance.

This study provides interesting insights into the complex dynamics between fiscal and monetary policy via their influence on the portfolio choices of different types of agents. It also suggests that the challenges for the conduct of monetary policy in the presence of substantial fiscal policy interventions, like in the last few years, are more complex than previously thought. Central Banks should also consider the impact of the ``self-insurance'' motive on the dynamics of the liquidity premium when setting the monetary policy stance.

Future work could extend the analysis presented here along several dimensions. For example, the model economy could be modified to allow for a CES production function with complementarity between capital and labor to check if the relative strength of the ``insurance-through-investment'' channel changes. Further extensions of the model should also consider the presence of distortionary taxes and targeted transfers. Finally, it would also be interesting to allow for endogenous regime switching in the fiscal policy mix.

\clearpage
\newpage
\bibliography{references}            
\bibliographystyle{natbib}     


\addcontentsline{toc}{section}{Appendix}

\setcounter{section}{0}
\setcounter{equation}{0}
\setcounter{figure}{0}
\setcounter{table}{0}
\numberwithin{equation}{section}
\numberwithin{figure}{section}
\numberwithin{table}{section}
\renewcommand{\theequation}{\Alph{section}\arabic{equation}}
\renewcommand{\thesection}{\Alph{section}\arabic{section}}
\renewcommand{\thesubsection}{\Alph{section}\arabic{subsection}}
\renewcommand{\thefigure}{\Alph{section}\arabic{figure}}
\renewcommand{\thetable}{\Alph{section}\arabic{table}}

\newpage
\appendix
\renewcommand{\thesection}{\Roman{section}}
\renewcommand{\thesubsection}{\Roman{section}.\roman{subsection}}
\part*{Appendix}

\section{Model details}\label{app:model_details}
\subsection{Solution}
The solution of the model is obtained by writing down the Bellman equations for capitalists and savers
\begin{align}
V^{K}(\mathbb{B}_{t}^{K}, K_{t}^{K}) &= \max_{\{N_t^K,Z_{t+1}, K_{t+1}^{K}\}_{t=0}^{\infty}}\left\{\frac{{C_{t}^{K}}^{1-\sigma}}{1-\sigma} - \psi\frac{{N_{t}^{K}}^{1+\varphi}}{1+\varphi} \right. \nonumber\\
&\left. +\beta\mathbb{E}_{t}\left[V^{K}(\mathbb{B}_{t+1}^{K}, K_{t+1}^{K})+\frac{\Pi_S}{\Pi_K}V^{S}(\mathbb{B}_{t+1}^{S})+\frac{\Pi_H}{\Pi_K}V\left(\mathbb{B}_{t+1}^{H}\right)\right]\right\} \nonumber\\ 
\text{s.t. }& \eqref{bc_k},\eqref{bc_u},\eqref{iac}, \eqref{lom_bk},\eqref{lom_bn},\eqref{lom_bu} \nonumber \\ 
\text{and similarly,} & \nonumber\\
V^{S}(\mathbb{B}_{t}^{S}) &= \max_{\{N_t^S,Z_{t+1}, \}_{t=0}^{\infty}}\left\{\frac{{C_{t}^{S}}^{1-\sigma}}{1-\sigma}- \psi\frac{{N_{t}^{S}}^{1+\varphi}}{1+\varphi}\right. \nonumber\\ 
&\left. +\beta\mathbb{E}_{t}\left[\frac{\Pi_K}{\Pi_S}V^{K}(\mathbb{B}_{t+1}^{K}, K_{t+1}^{K})+V^{S}(\mathbb{B}_{t+1}^{S})+\frac{\Pi_H}{\Pi_S}V\left(\mathbb{B}_{t+1}^{H}\right)\right]\right\} \nonumber\\ 
\text{s.t. }& \eqref{bc_n},\eqref{bc_u},\eqref{lom_bk},\eqref{lom_bn},\eqref{lom_bu}. \nonumber
\end{align}
Moreover, hand-to-mouth agents will not consume on their Euler equation, but will choose the amount of labor hours optimally, giving rise to a standard intra-temporal condition detailed below.

Using dynamic programming techniques, we can show that the optimality conditions to these programs can be expressed in terms of three Euler equations,
\begin{align}
    {C_{t}^{K}}^{-\sigma} & = \beta\mathbb{E}_{t}\left[\frac{\left(R_{t+1}^{K}+(1-\delta)Q_{t+1}\right){C_{t+1}^{K}}^{-\sigma}}{Q_{t}}\right] \label{eu_k} \\
    {C_{t}^{K}}^{-\sigma} & = \beta\mathbb{E}_{t}\left[\frac{1+R_{t}^{B}}{\Pi_{t+1}}\left(\lambda_{K,K}{C_{t+1}^{K}}^{-\sigma}+\lambda_{K,S}{C_{t+1}^{S}}^{-\sigma}+\lambda_{K,H}{C_{t+1}^{H}}^{-\sigma}\right)\right] \label{eu_bk} \\
    {C_{t}^{S}}^{-\sigma} & = \beta\mathbb{E}_{t}\left[\frac{1+R_{t}^{B}}{\Pi_{t+1}}\left(\lambda_{S,K}{C_{t+1}^{K}}^{-\sigma}+\lambda_{S,S}{C_{t+1}^{S}}^{-\sigma}+\lambda_{S,H}{C_{t+1}^{H}}^{-\sigma}\right)\right], \label{eu_bn}
\end{align}

\todo[inline]{$R^K$ in the code is in gross terms. Like all other interest rates.}
and three intra-temporal conditions 
\begin{align}
    \psi\frac{{N_{t}^{K}}^{\varphi}}{{C_{t}^{K}}^{-\sigma}} &= W_t \label{intra_k}\\
    \psi\frac{{N_{t}^{S}}^{\varphi}}{{C_{t}^{S}}^{-\sigma}} &= W_t \label{intra_n}\\ 
    \psi\frac{{N_{t}^{H}}^{\varphi}}{{C_{t}^{H}}^{-\sigma}} &= W_t .\label{intra_u}
\end{align}
\subsection{labor union}
Following much of the literature on models with limited heterogeneity, we assume the existence of a labor union that pools the labor supplies of the different types, sets the wage and redistributes labor hours equally among agents ($N^K_t=N^S_t=N^H_t$). Hence the total labor supply in the economy is given by 
\begin{equation}
\psi N_t^{\varphi} = W_t{C_{t}}^{-\sigma},
\end{equation}
where $\psi$ can be calibrated in order to ensure that the number of aggregate hours worked in steady state is $0.33$.

\subsection{Full list of equilibrium conditions}

\begin{equation}
    {U^{\prime}(C_{t}^K)}= {C_{t}^K}^{-{{\sigma}}}
    \end{equation}
    \begin{equation}
    {U^{\prime}(C_{t}^K)}={{\beta}}\, {R}_{t}\, \left({{\lambda_{K,K}}}\, {U^{\prime}(C_{t+1}^K)}+{{\lambda_{K,S}}}\, {U^{\prime}(C_{t+1}^S)}+{{\lambda_{K,H}}}\, {U^{\prime}(C_{t+1}^H)}\right)
    \end{equation}
    \begin{equation}
    {U^{\prime}(N_{t}^K)}=-{{\psi}}\, {N_{t}^K}^{{{\varphi}}}
    \end{equation}
    \begin{equation}
    {W}_{t}={{\psi}}\, {N}_{t}^{{{\varphi}}} {C}_{t}^{{{\sigma}}}
    \end{equation}
    \begin{equation}
    {N_{t}^K}={N}_{t}
    \end{equation}
    \begin{equation}
    {\mathbb{B}_{t}^{K}}={{\lambda_{K,K}}}\, {{\Pi_{K}}}\, {Z_{t}^{K}}+{{\Pi_{S}}}\, {{\lambda_{S,K}}}\, {Z_{t}^{S}}
    \end{equation}
    \begin{equation}
    {U^{\prime}(C_{t}^S)}= {C_{t}^S}^{-{{\sigma}}}
    \end{equation}
    \begin{equation}
    {U^{\prime}(C_{t}^S)}={{\beta}}\, {R}_{t}\, \left({U^{\prime}(C_{t+1}^S)}\, {{\lambda_{S,S}}}+{U^{\prime}(C_{t+1}^K)}\, {{\lambda_{S,K}}}+{U^{\prime}(C_{t+1}^H)}\, {{\lambda_{S,H}}}\right)
    \end{equation}
    \begin{equation}
    {U^{\prime}(N_{t}^S)}=-{{\psi}}\, {N_{t}^S}^{{{\varphi}}}
    \end{equation}
    \begin{equation}
    {N_{t}^S}={N}_{t}
    \end{equation}
    \begin{equation}
    {C_{t}^S}+\frac{{Z_{t+1}^{S}}}{{{P}}_{t}}={W}_{t}\, {N_{t}^S}+\frac{(1+{R_{t-1}^b})\, {\mathbb{B}_{t}^{S}}}{{{\Pi_{S}}}\, {{P}}_{t}}-\tau^S-\frac{{T_{t}^S}}{{{P}}_{t}}
    \end{equation}
    \begin{equation}
    {\mathbb{B}_{t}^{S}}={Z_{t}^{S}}\, {{\Pi_{S}}}\, {{\lambda_{S,S}}}+{Z_{t}^{K}}\, {{\lambda_{K,S}}}\, {{\Pi_{K}}}
    \end{equation}
    \begin{equation}
    {U^{\prime}(C_{t}^H)}= {C_{t}^H}^{-{{\sigma}}}
    \end{equation}
    \begin{equation}
    {U^{\prime}(C_{t}^H)}=-{{\psi}}\, {N_{t}^H}^{{{\varphi}}}
    \end{equation}
    \begin{equation}
    {N_{t}^H}={N}_{t}
    \end{equation}
    \begin{equation}
    {Z_{t}^{H}}=0
    \end{equation}
    \begin{equation}
    {\mathbb{B}_{t}^{H}}={Z^{K}_{t}}\, {{\lambda_{K,H}}}\, {{\Pi_{K}}}+{Z_{t}^{S}}\, {{\Pi_{S}}}\, {{\lambda_{S,H}}}
    \end{equation}
    \begin{equation}
    {C_{t}^H}={W}_{t}\, {N^H}_{t}+(1+{R_{t-1}^b})\, \frac{{\mathbb{B}_{t}^{H}}}{{{\Pi_{H}}}\, {{P}}_{t}}-\frac{{T_{t}^H}}{{{P}}_{t}}+\tau^H
    \end{equation}
    \begin{equation}
    {{Y}}_{t}=\left({N}_{t}\, {A}_{t}\right)^{1-{{\alpha}}}\, {{K}}_{t-1}^{{{\alpha}}}-F
    \end{equation}
    \begin{equation}
    {MPL}_{t}=\left(1-{{\alpha}}\right)\, {A}_{t}^{1-{{\alpha}}}\, \left(\frac{{N}_{t}}{{{K}}_{t-1}}\right)^{-{{\alpha}}}
    \end{equation}
    \begin{equation}
    {W}_{t}={MPL}_{t}\, {MC}_{t}
    \end{equation}
    \begin{equation}
    {Y}_{t}={D}_{t}+{N}_{t}\, {W}_{t}+{{R_{t}^K}}\, {{K}}_{t-1}
    \end{equation}
    \begin{equation}
    {{MPK}}_{t}={{\alpha}}\, \left(\frac{{{K}}_{t-1}}{{N}_{t}\, {A}_{t}}\right)^{{{\alpha}}-1}
    \end{equation}
    \begin{equation}
    {{R_{t}^K}}={MC}_{t}\, {{MPK_{t}}}
    \end{equation}
    \begin{equation}
    {{K_{t}^K}}={{I_{t}^K}}\, \left(1-{{S}}_{t}\right)\, +{{K_{t-1}^K}}\, \left(1-{{\delta}}\right)
    \end{equation}
    \begin{equation}
    {{X}}_{t}=\frac{{{I_{t}^K}}}{{{I_{t-1}^K}}}
    \end{equation}
    \begin{equation}
    {{S}}_{t}={{\iota}}\, \left({{X}}_{t}-1\right)^{2}
    \end{equation}
    \begin{equation}
    {{S_{t}^{\prime}}}=\left({{X}}_{t}-1\right)\, 2\, {{\iota}}
    \end{equation}
    \begin{equation}
    {U^{\prime}(C_{t}^K)}={U^{\prime}(C_{t+1}^K)}\, \frac{{{\beta}}\, \left({{R_{t+1}^K}}+\left(1-{{\delta}}\right)\, {{Q}}_{t+1}\right)}{{{Q}}_{t}}
    \end{equation}
    \begin{equation}
    {{LP}}_{t}=\frac{1}{{R}_{t}}\, \frac{{{R_{t+1}^K}}+\left(1-{{\delta}}\right)\, {{Q}}_{t+1}}{{{Q}}_{t}}
    \end{equation}
    \begin{equation}
    {{Q}}_{t}\, \left(1-{{S}}_{t}-{{X}}_{t}\, {{S_{t}^{\prime}}}\right)+{{Q}}_{t+1}\, \frac{{U^{\prime}(C_{t+1}^K)}\, {{\beta}}\,}{{U^{\prime}(C_{t}^K)}}\, {{S_{t+1}^{\prime}}}\, {{X}}_{t+1}^{2}=1
    \end{equation}
    \begin{equation}
    {{I}}_{t}={{\Pi_{K}}}\, {{I_{t}^K}}
    \end{equation}
    \begin{equation}
    {{K}}_{t}={{\Pi_{K}}}\, {{K_{t}^K}}
    \end{equation}
    \begin{equation}
    {Y}_{t}={{I}}_{t}+{C}_{t}+\frac{{G}_{t}}{{{P}}_{t}}
    \end{equation}
    \begin{equation}
    {R}_{t}=\frac{(1+{R_{t}^b})}{{\Pi}_{t+1}}
    \end{equation}
    \begin{equation}
    {C}_{t}={{\Pi_{H}}}\, {C_{t}^H}+{C_{t}^K}\, {{\Pi_{K}}}+{{\Pi_{S}}}\, {C_{t}^S}
    \end{equation}
    \begin{equation}
    1-{{\epsilon}}+{MC}_{t}\, {{\epsilon}}\, {{MS}}_{t}-{\Pi}_{t}\, {{\xi}}\, \left({\Pi}_{t}-1\right)+{{\beta}}\, \frac{{{\xi}}\, {\Pi}_{t+1}\, \frac{{U^{\prime}(C_{t+1}^K)}}{{U^{\prime}(C_{t}^K)}}\, \left({\Pi}_{t+1}-1\right)\, {Y}_{t+1}}{{Y}_{t}}=0
    \end{equation}
    \begin{equation}
    \frac{1 + R^b_{t}}{1+R_b^{*}} = \left(\frac{1 + R^b_{t-1}}{1+R_b^{*}}\right)^{\rho_{R^b}}\left(\frac{\Pi_{t}}{\Pi^{*}}\right)^{(1-\rho_{R^b})\phi_{\pi}}M_t,
    \end{equation}
    \begin{equation}
    \frac{M_t}{M^{*}}=\left(\frac{M_{t-1}}{M^{*}}\right)^{\rho_m}\epsilon_{m,t},
    \end{equation}
    \begin{equation}
    \log\left({A}_{t}\right)-\log\left((A^*)\right)={{\rho_{a}}}\, \left(\log\left({A}_{t-1}\right)-\log\left((A^*)\right)\right)+{{\epsilon_{t}^{A}}}
    \end{equation}
    \begin{equation}
    {B}_{t+1}={G}_{t}+(1+{R^b_{t-1}})\, {B}_{t}-{T}_{t}
    \end{equation}
    \begin{equation}
    {B}_{t}={{\Pi_{K}}}\, {Z_{t}^{K}}+{{\Pi_{S}}}\, {Z_{t}^{S}}
    \end{equation}
    \begin{equation}
    {\Pi}_{t}=\frac{{{P}}_{t}}{{{P}}_{t-1}}
    \end{equation}
    \begin{equation}
    {S^b}_{t}=\frac{{B}_{t}}{{{P}}_{t}\, {Y}_{t}}
    \end{equation}
    \begin{equation}
    \frac{G_t}{G^{*}}=\left(\frac{G_{t-1}}{G^{*}}\right)^{\rho_g}\epsilon_{g,t}
    \end{equation}
    \begin{equation}
    \frac{T_t}{T^{*}}=\left(\frac{T_{t-1}}{T^{*}} \right)^{\rho_T } \left(\frac{B_{t-1}}{B^{*}} \right)^{(1-\rho_T)\gamma_T}\left(\frac{G_{t}}{G^{*}} \right)^{(1-\rho_T)\gamma_{TG}}\epsilon_{t,t}
    \end{equation}
    \begin{equation}
    {T_{t}^K}={T_{t}^H}={T_{t}^S}= {T}_{t}
    \end{equation}

\subsection{Steady State}

\begin{eqnarray*}
    P^*=\Pi^*\\
    R^*={R^b}^*=\frac{1}{\beta}-1\\
    MC^*=\frac{\epsilon}{\epsilon-1}\\
    {R^K}^*=(R^*-1+\delta)\\
    {N^K}^*={N^S}^*={N^H}^*=N^*\\
    K^*=\left(\frac{{R^K}^*}{(MC^* \alpha)}\right)^\frac{1}{\alpha-1} N^*\\
    W^*=MC^*(1-\alpha)\left(\frac{N^*}{K^*}\right)^{-\alpha}\\
    F=N^* \left(\left(\frac{N^*}{K^*}\right)^{-\alpha}-\left(W^*+{R^K}^* \frac{K^*}{N^*}\right)\right)\\
    Y={N^*}^(1-\alpha){K^*}^{\alpha}-F\\
    B^*=0.57Y^*4\\
    I^*=\delta K^*\\
    G^*=0.2Y^*\\
    C^*=Y^*-I^*-G^*\\
    {C^K}^*={C^S}^*={C^H}^*=C^*\\
    T^*=R^*B^*+G^*-B^*\\
    {T^K}^*={T^H}^*={T^S}^*= {T}^*\\
    {Z^H}^*={Z^S}^*=0\\
    {Z^K}^*=\frac{B^*}{\Pi_K}
\end{eqnarray*}

\clearpage
\newpage

\subsection{Proofs} \label{proofs}

\textbf{Proof of Proposition \ref{prop_1}}

\begin{proof}
Consider the two Euler equations for the capitalist type,
\begin{align}
    {C_{t}^{K}}^{-\sigma} & = \beta\mathbb{E}_{t}\left[\mathcal{R}_{t+1}^{K}{C_{t+1}^{K}}^{-\sigma}\right] \label{eu_k} \\
    {C_{t}^{K}}^{-\sigma} & = \beta\mathbb{E}_{t}\left[\mathcal{R}_{t+1}^{B}\left(\lambda_{KK}{C_{t+1}^{K}}^{-\sigma}+\lambda_{KS}{C_{t+1}^{S}}^{-\sigma}+\lambda_{KH}{C_{t+1}^{H}}^{-\sigma}\right)\right] \label{eu_bk} 
\end{align}
where we have now used the definitions of the gross real return on capital $\left(\mathcal{R}_{t+1}^{K}=\frac{\left(R_{t+1}^{K}+(1-\delta)Q_{t+1}\right)}{Q_t}\right)$ and of the gross real return on bonds $\left(\mathcal{R}_{t+1}^{B}=\frac{1+R_{t}^{B}}{\Pi_{t+1}}\right)$ respectively.\todo{All nominal interest rates should be expressed in gross terms throughout.}

Then, by taking a first-order approximation of the two around the deterministic steady state and combining them we obtain
\begin{equation}
    \beta\mathbb{E}_{t}\left[\Tilde{\mathcal{R}}^{K}_{t+1} - \Tilde{\mathcal{R}}^{B}_{t+1}\right] = -\sigma\mathbb{E}_{t}\left[\lambda_{KS}(\hat{C}^S_{t+1}-\hat{C}^K_{t+1})+\lambda_{KH}(\hat{C}^H_{t+1}-\hat{C}^K_{t+1})\right] \label{lp_step_2}
\end{equation}

where ``$\Tilde{(\cdot)}$'' denotes that the variable has been linearized, as opposed to log-linearized (denoted by `$\hat{(\cdot)}$''), in line with how the literature treats variables that are already expressed as percentages. 

Finally, we conclude the proof by recognizing that the left-hand side of equation \eqref{lp_step_2} is the approximation to the first order of the liquidity premium as defined in the main text.
\end{proof}

\clearpage
\newpage
\appendix
\part*{Online Appendix - Not for Publication}
\section{Lump sum transfer shock}\label{sec:Tshock}
In this section we show results to a temporary decline in $\epsilon_{t,t}$ in the fiscal rule for lump sum taxes. This corresponds to a temporary increase in lump sum transfers. Results are in line with the analysis in the manuscript.

\begin{figure}[h!]
    \centering
    \includegraphics[width=\textwidth]{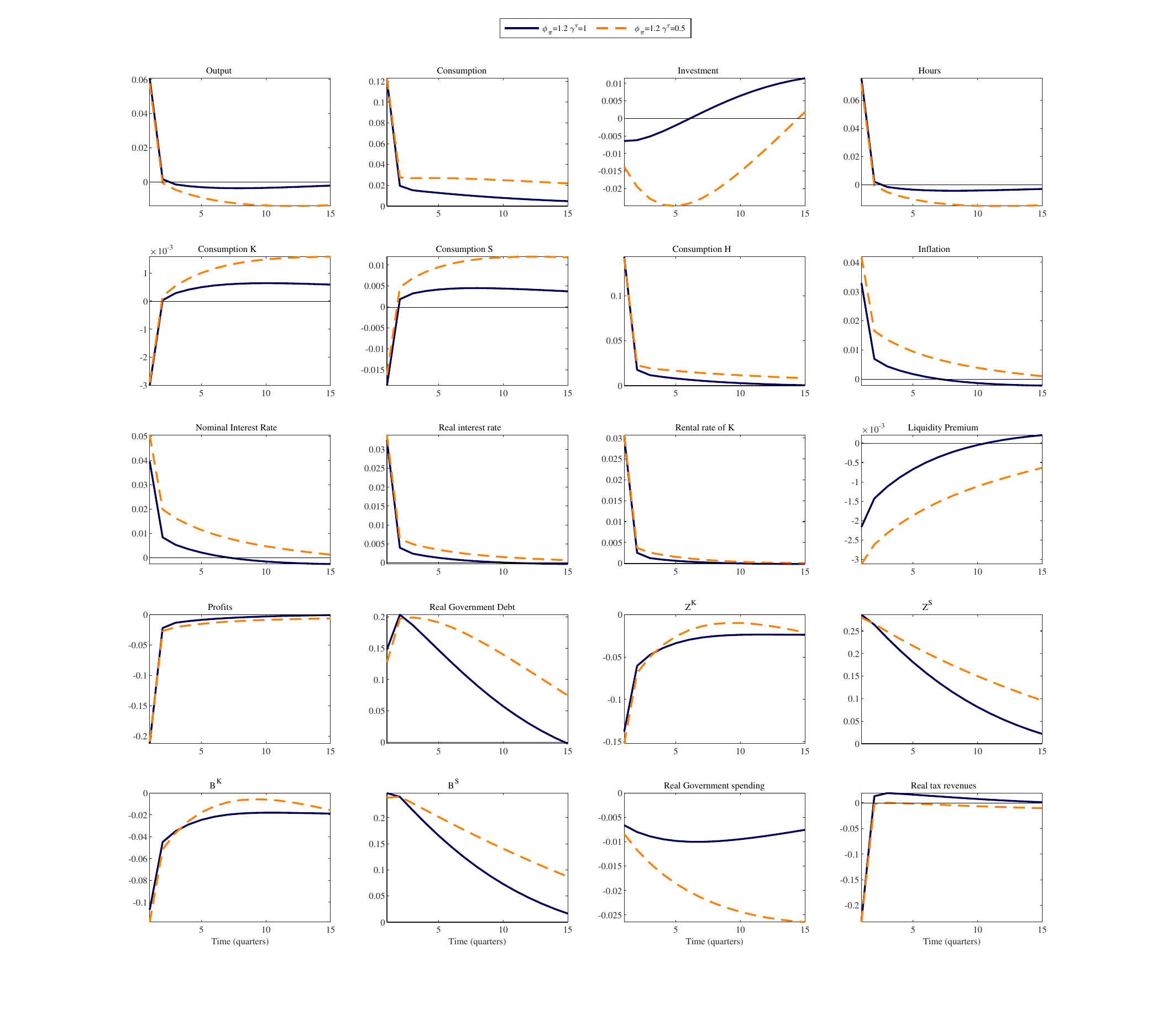}
    \caption{Impulse responses to a temporary 1\% decline in $\epsilon_{t,t}$ for the two fiscal policy regimes when monetary policy is active ($\phi_{\pi}=1.2$).\\
    \begin{tiny}
        Note: All variables are expressed in real terms except for Hours, Inflation, and Nominal interest rate. All variables related to fiscal policy are in \% deviation from the steady state of output. The remaining variables are in \% deviations from their steady state. We plot the next period realized rental rate of capital ($R^K_{t+1}$). Consumption, Z's, and B's are island-wide figures (multiplied by the population sizes $\Pi$'s).
    \end{tiny}}
    \label{fig:3A_AM_epsT}
    \end{figure}
    
    \begin{figure}[h!]
        \centering
        \includegraphics[width=\textwidth]{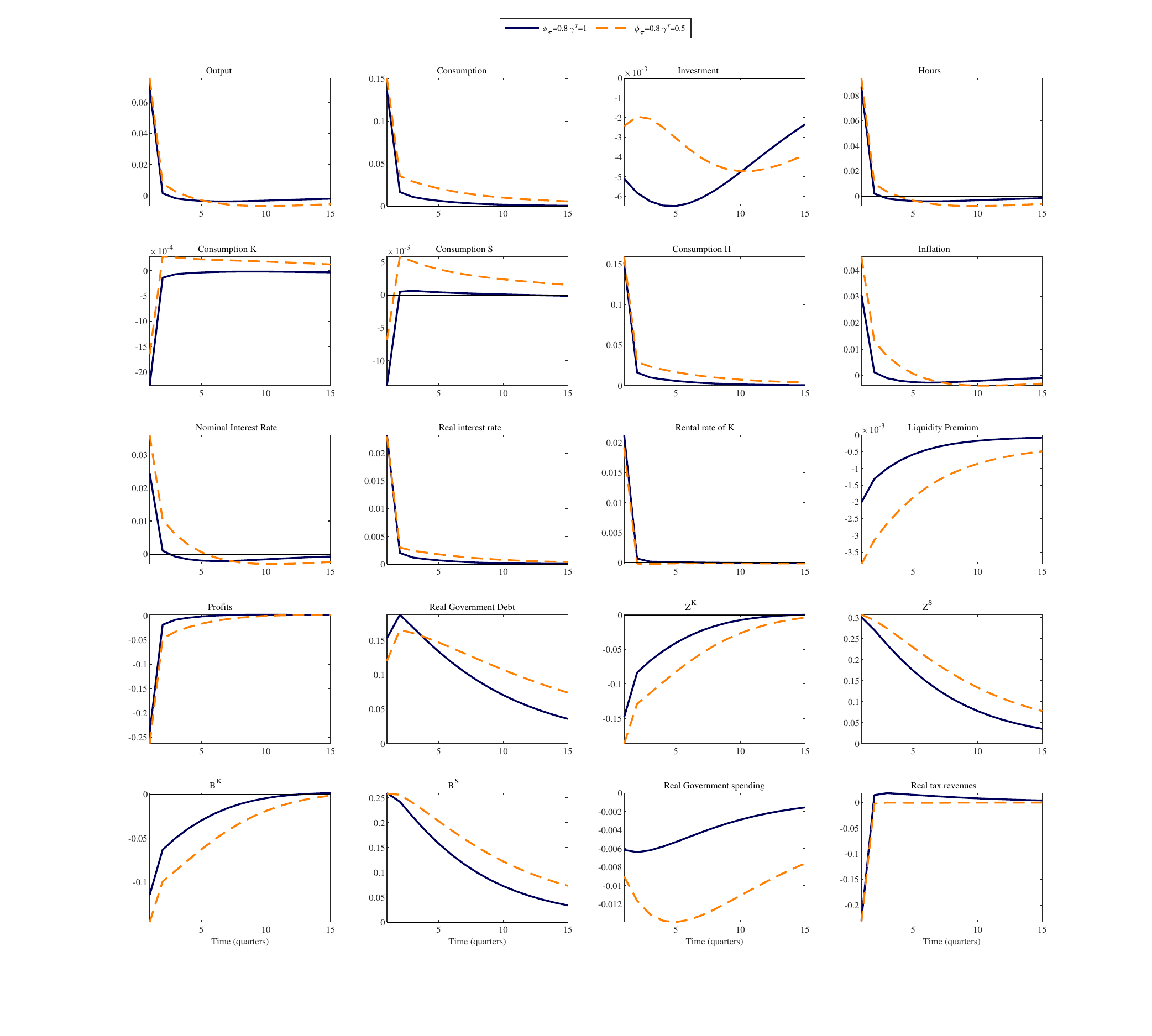}
        \caption{Impulse responses to a temporary 1\% decline in $\epsilon_{t,t}$ for the two fiscal policy regimes when monetary policy is passive ($\phi_{\pi}=0.8$).\\
            \begin{tiny}
            Note: All variables are expressed in real terms except for Hours, Inflation, and Nominal interest rate. All variables related to fiscal policy are in \% deviation from the steady state of output. The remaining variables are in \% deviations from their steady state. We plot the next period realized rental rate of capital ($R^K_{t+1}$). Consumption, Z's, and B's are island-wide figures (multiplied by the population sizes $\Pi$'s).
    \end{tiny}}
        \label{fig:3A_PM_epsT}
    \end{figure}

\clearpage
\newpage
    
\section{Robustness}\label{sec:Robustness}

In this section, we show the sensitivity of our results to the calibration of the transition probabilities. As equation \eqref{liquidity_premium} shows, the key probabilities driving the liquidity premium are $\lambda_{K,S}$ and $\lambda_{K,H}$. Hence, we focus on the calibration of the probabilities related to the flows out of the capitalists island. Precisely we compare the baseline calibration ($\lambda_{K,K}=0.8$, $\lambda_{K,H}=0.4$) against a higher and lower probability of remaining capitalists (0.9 and 0.7 respectively) and a higher probability of becoming hand-to-mouth (0.05). We show results for the fiscal shock\footnote{Technology and monetary policy shocks results show similar patterns and are available upon request.} and present the possible combinations between fiscal and monetary policies separately in this case for ease of exposition.
The next four figures show that, while there are obvious quantitative differences in the impulse response functions, the overall qualitative picture is not affected by the transition probabilities.

\begin{figure}[h!]
    \centering
    \includegraphics[width=\textwidth]{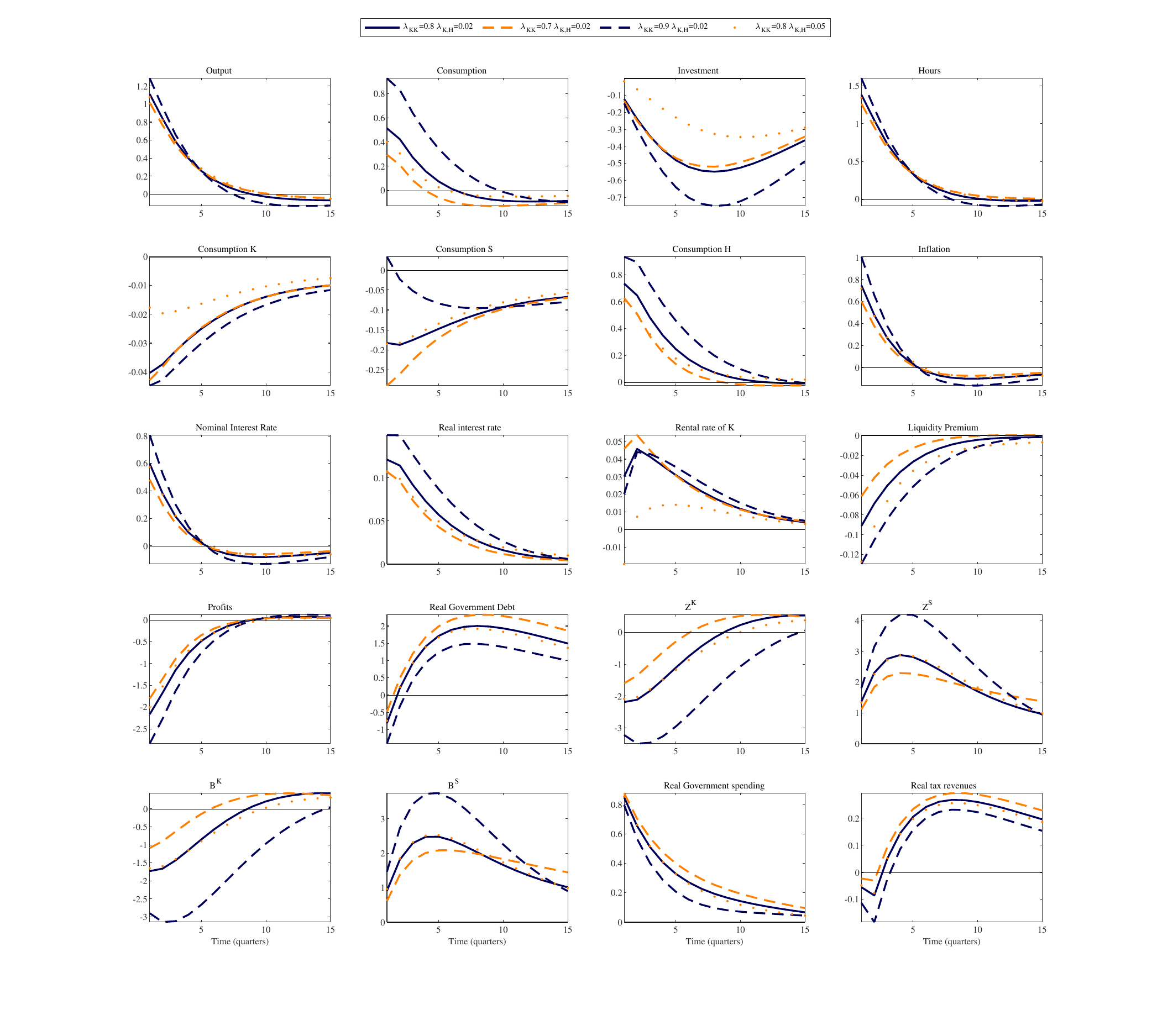}
    \caption{Impulse responses to a temporary 1\% increase in G with $\gamma_T=1$ and $\phi_{\pi}=0.8$.\\
    \begin{tiny}
        Note: All variables are expressed in real terms except for Hours, Inflation, and Nominal interest rate. All variables related to fiscal policy are in \% deviation from the steady state of output. The remaining variables are in \% deviations from their steady state. We plot the next period realized rental rate of capital ($R^K_{t+1}$). Consumption, Z's, and B's are island-wide figures (multiplied by the population sizes $\Pi$'s).
    \end{tiny}}
    \end{figure}

    \begin{figure}[h!]
        \centering
        \includegraphics[width=\textwidth]{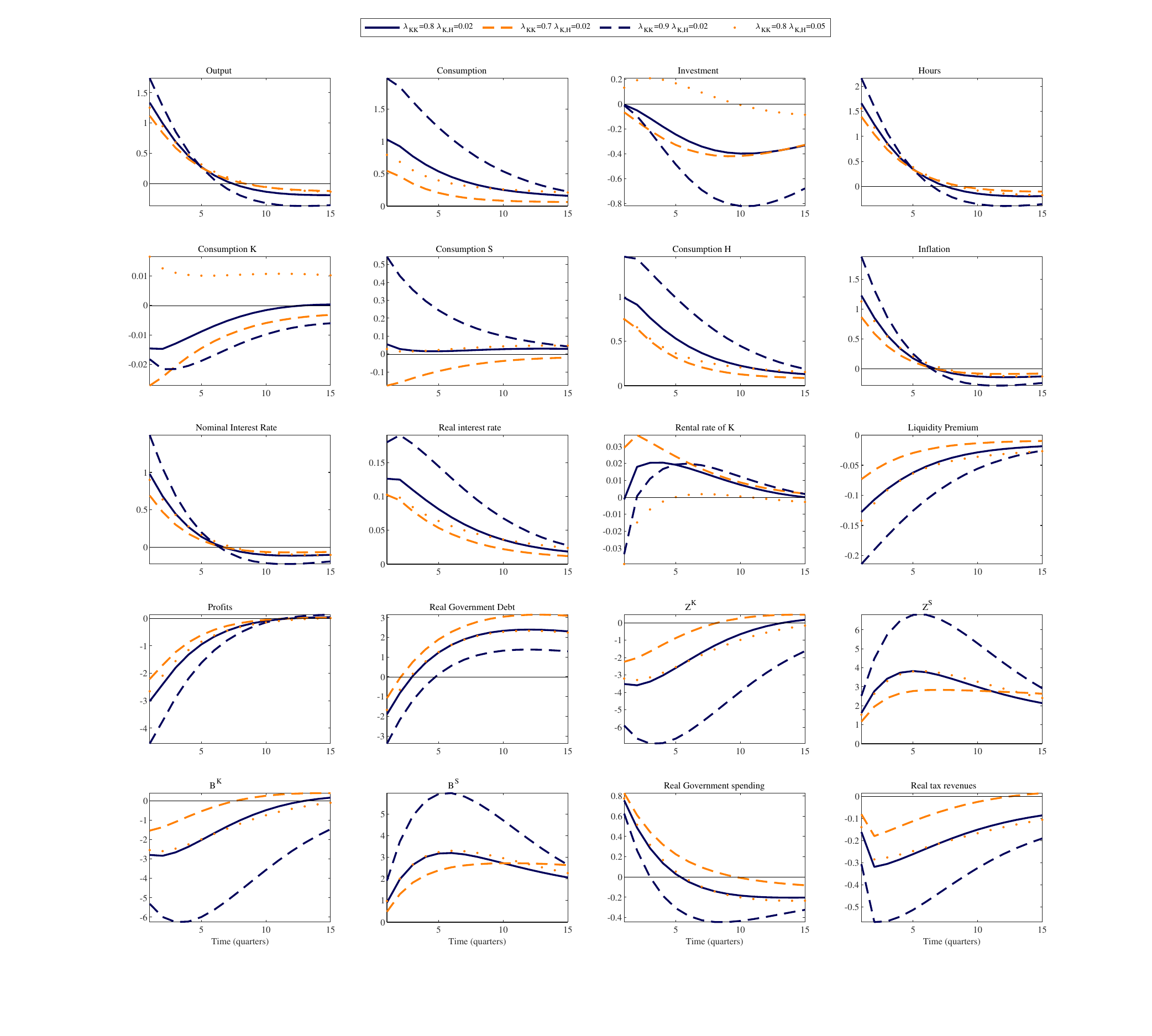}
        \caption{Impulse responses to a temporary 1\% increase in G with $\gamma_T=0.5$ and $\phi_{\pi}=0.8$.\\
        \begin{tiny}
            Note: All variables are expressed in real terms except for Hours, Inflation, and Nominal interest rate. All variables related to fiscal policy are in \% deviation from the steady state of output. The remaining variables are in \% deviations from their steady state. We plot the next period realized rental rate of capital ($R^K_{t+1}$). Consumption, Z's, and B's are island-wide figures (multiplied by the population sizes $\Pi$'s).
        \end{tiny}}
        \end{figure}

        \begin{figure}[h!]
            \centering
            \includegraphics[width=\textwidth]{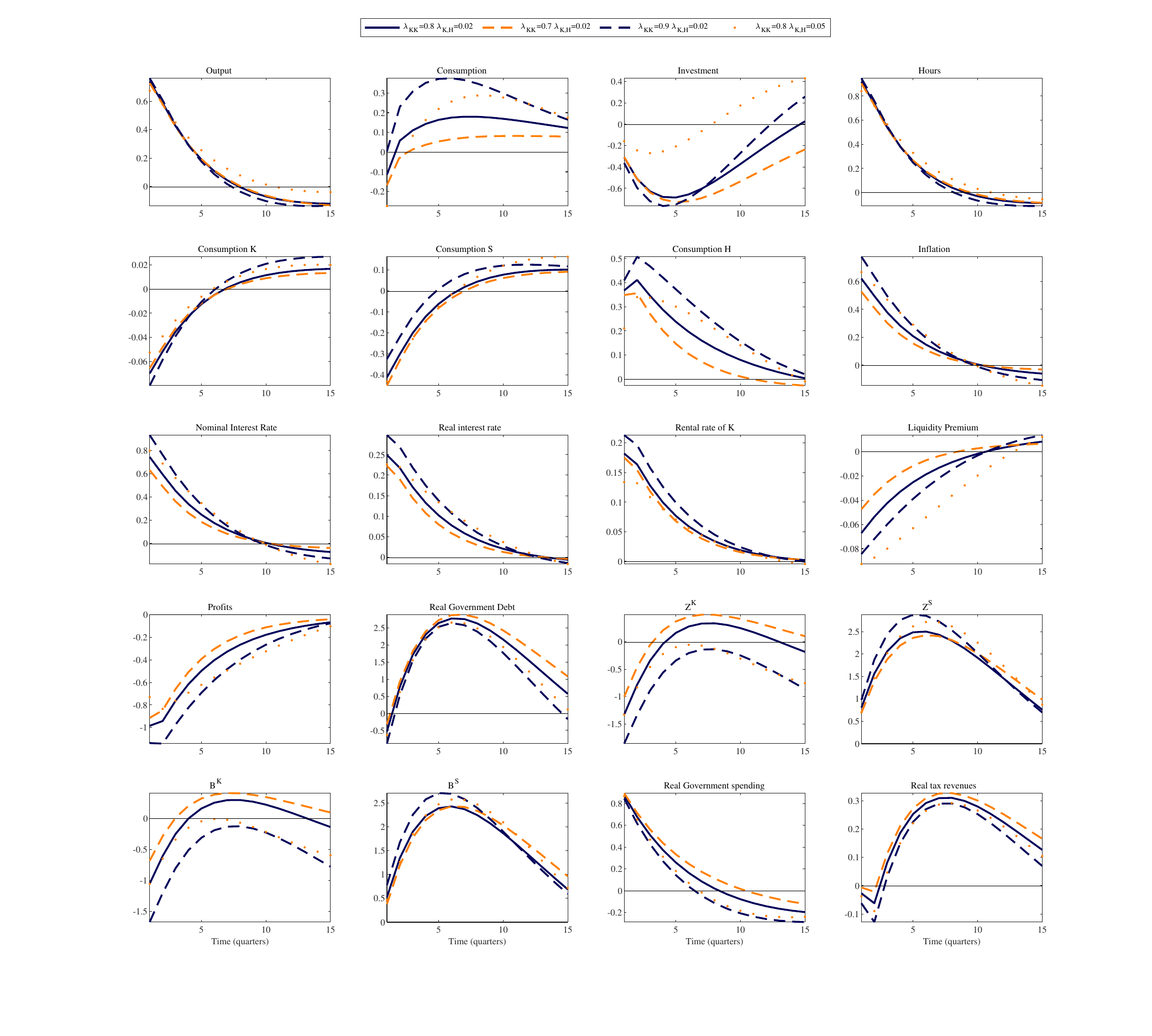}
            \caption{Impulse responses to a temporary 1\% increase in G with $\gamma_T=1$ and $\phi_{\pi}=1.2$.\\
            \begin{tiny}
                Note: All variables are expressed in real terms except for Hours, Inflation, and Nominal interest rate. All variables related to fiscal policy are in \% deviation from the steady state of output. The remaining variables are in \% deviations from their steady state. We plot the next period realized rental rate of capital ($R^K_{t+1}$). Consumption, Z's, and B's are island-wide figures (multiplied by the population sizes $\Pi$'s).
            \end{tiny}}
            \end{figure}

            \begin{figure}[h!]
                \centering
                \includegraphics[width=\textwidth]{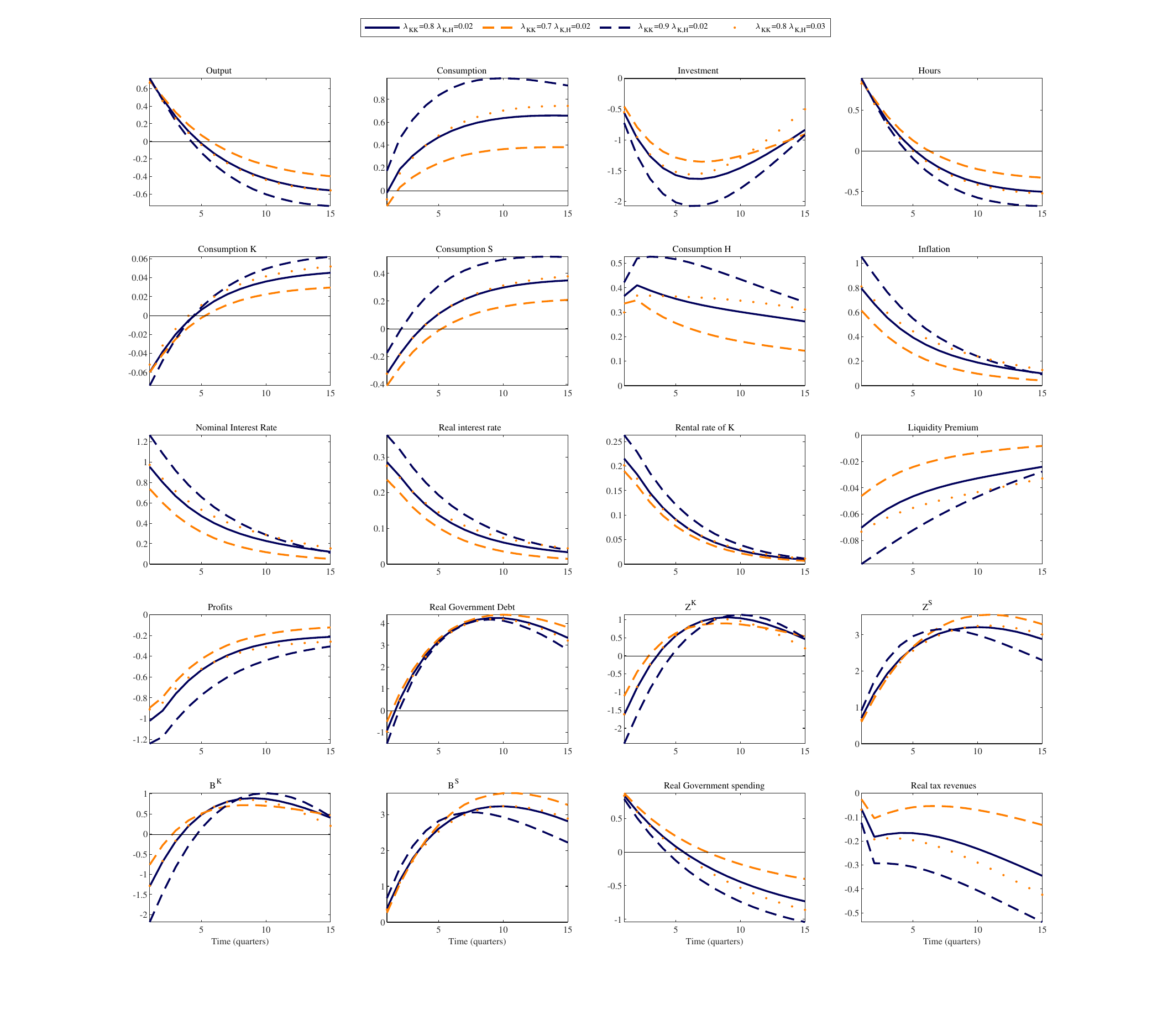}
                \caption{Impulse responses to a temporary 1\% increase in G with $\gamma_T=0.5$ and $\phi_{\pi}=1.2$.\\
                \begin{tiny}
                    Note: All variables are expressed in real terms except for Hours, Inflation, and Nominal interest rate. All variables related to fiscal policy are in \% deviation from the steady state of output. The remaining variables are in \% deviations from their steady state. We plot the next period realized rental rate of capital ($R^K_{t+1}$). Consumption, Z's, and B's are island-wide figures (multiplied by the population sizes $\Pi$'s). Here the orange dotted line is produced for a value of $\lambda_{K,H}=0.03$ instead. Higher values would lead to multiple equilibria.
                \end{tiny}}
                \end{figure}

\clearpage
\newpage
    
\section{Stability analysis}\label{sec:Stability}

\begin{figure}[h!]
   \caption{Stability region for different values of $\phi_{\Pi}$, and $\Pi_H$ in the two-agent model. }
   \label{fig:stab}
    \begin{subfigure}[b]{0.45\textwidth}
        \centering
        \includegraphics[width=\textwidth]{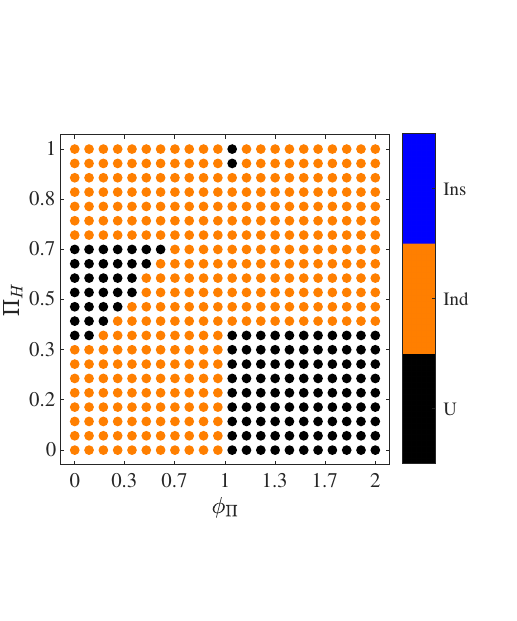}
        \caption{Real FP}
    \end{subfigure}
    \hfill
    \begin{subfigure}[b]{0.45\textwidth}
        \centering
        \includegraphics[width=\textwidth]{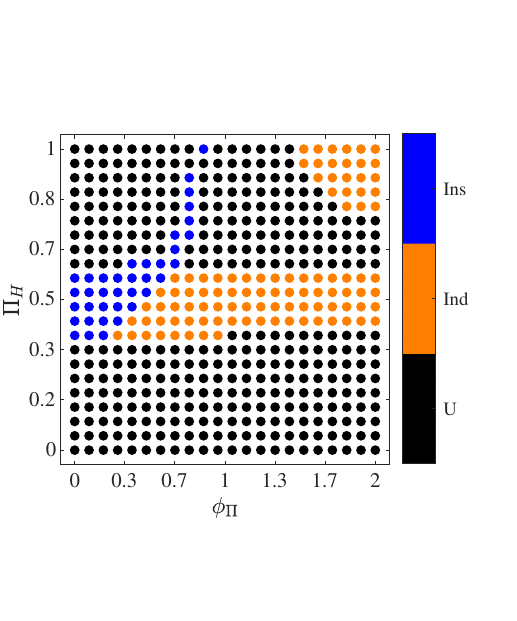}
        \caption{Nominal FP}
    \end{subfigure}
    \hfill
\subcaption*{\textit{Notes: Real FP is the standard model where fiscal policy is specified in real terms. Nominal FP is the model with nominal fiscal policy.} }
\end{figure}

\begin{figure}[h!]
   \caption{Stability region for different values of $\phi_{\Pi}$, and $\Pi_H$ in the three-agent model. }
   \label{fig:stab}
    \begin{subfigure}[b]{0.45\textwidth}
        \centering
        \includegraphics[width=\textwidth]{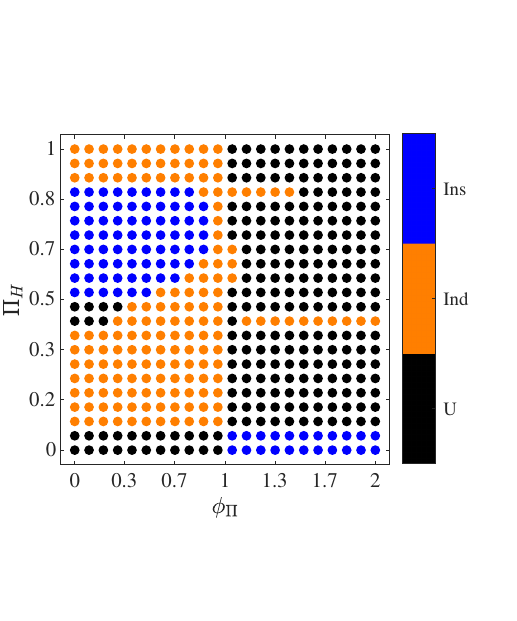}
        \caption{Real FP}
    \end{subfigure}
    \hfill
    \begin{subfigure}[b]{0.45\textwidth}
        \centering
        \includegraphics[width=\textwidth]{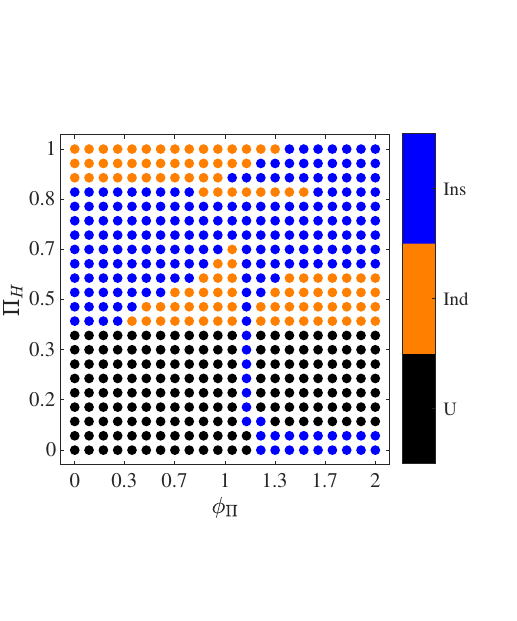}
        \caption{Nominal FP}
    \end{subfigure}
    \hfill
\subcaption*{\textit{Notes: Real FP is the standard model where fiscal policy is specified in real terms. Nominal FP is the model with nominal fiscal policy.} }
\end{figure}

\begin{figure}
   \caption{Stability region for different values of $\phi_{\Pi}$, and $\gamma_T$ in the two-agent model. }
   \label{fig:stab}
    \begin{subfigure}[b]{0.45\textwidth}
        \centering
        \includegraphics[width=\textwidth]{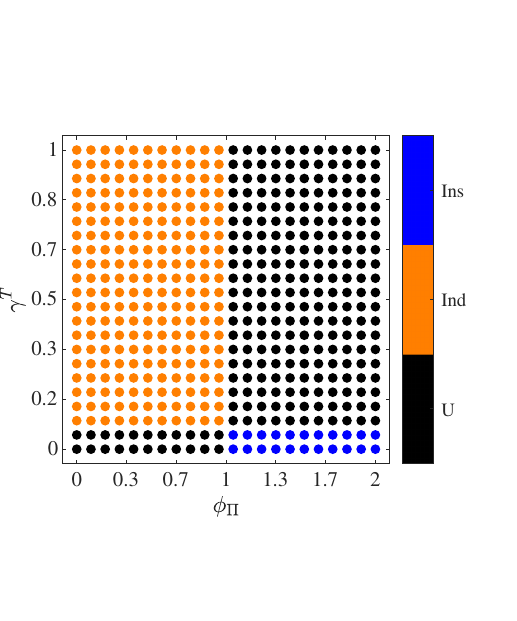}
        \caption{Real FP}
    \end{subfigure}
    \hfill
    \begin{subfigure}[b]{0.45\textwidth}
        \centering
        \includegraphics[width=\textwidth]{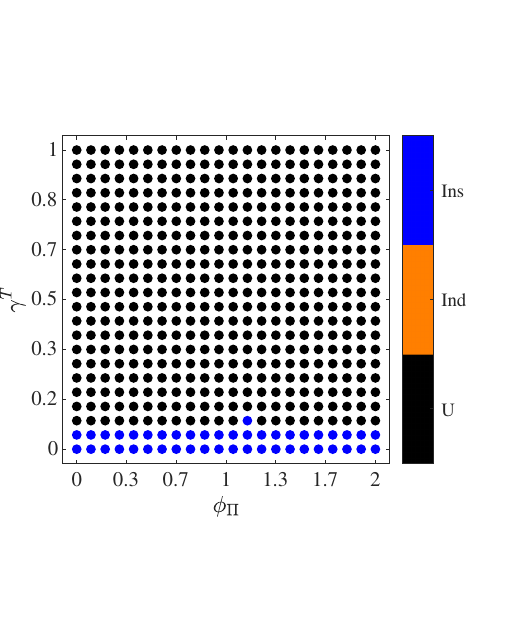}
        \caption{Nominal FP}
    \end{subfigure}
    \hfill
\subcaption*{\textit{Notes: Real FP is the standard model where fiscal policy is specified in real terms. Nominal FP is the model with nominal fiscal policy.} }
\end{figure}

\begin{figure}[h!]
   \caption{Stability region for different values of $\phi_{\Pi}$, and $\gamma_T$ in the three-agent model. }
   \label{fig:stab}
    \begin{subfigure}[b]{0.45\textwidth}
        \centering
        \includegraphics[width=\textwidth]{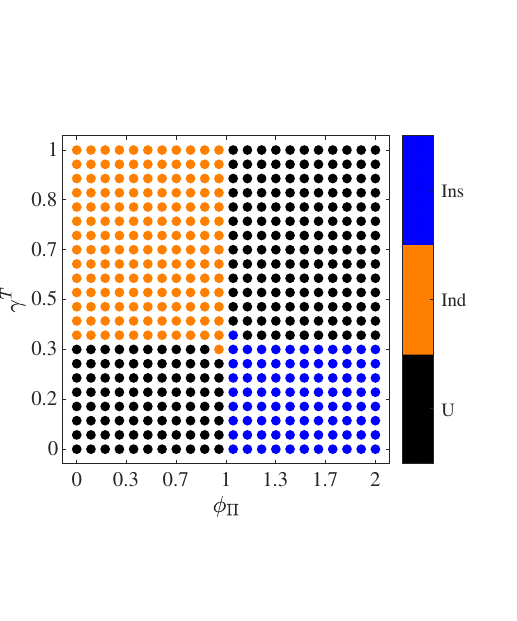}
        \caption{Real FP}
    \end{subfigure}
    \hfill
    \begin{subfigure}[b]{0.45\textwidth}
        \centering
        \includegraphics[width=\textwidth]{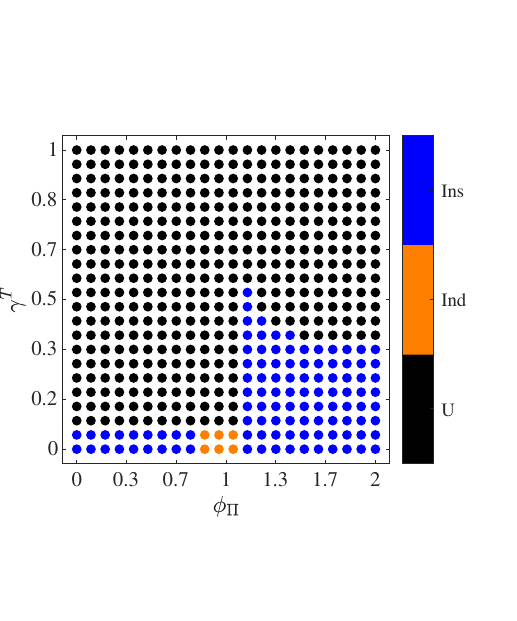}
        \caption{Nominal FP}
    \end{subfigure}
    \hfill
\subcaption*{\textit{Notes: Real FP is the standard model where fiscal policy is specified in real terms. Nominal FP is the model with nominal fiscal policy.} }
\end{figure}

\clearpage
\newpage
    
\section{Comparison with the two-agent model}\label{sec:2A}
\todo[inline]{Why no differences in aggregates?}
In this section, we report the responses to each business cycle shock considered in the main text when we shut down the self-insurance mechanism by assuming only two agents as in \cite{Bilbiie:21} and \cite{bilbiie2019capital}. This means that we remove the savers island $\Pi_S=0$ and we set $\Pi_K=0.8$  and $\lambda_{K,K}=0.98$.


    \begin{figure}[h!]
        \centering
        \includegraphics[width=\textwidth]{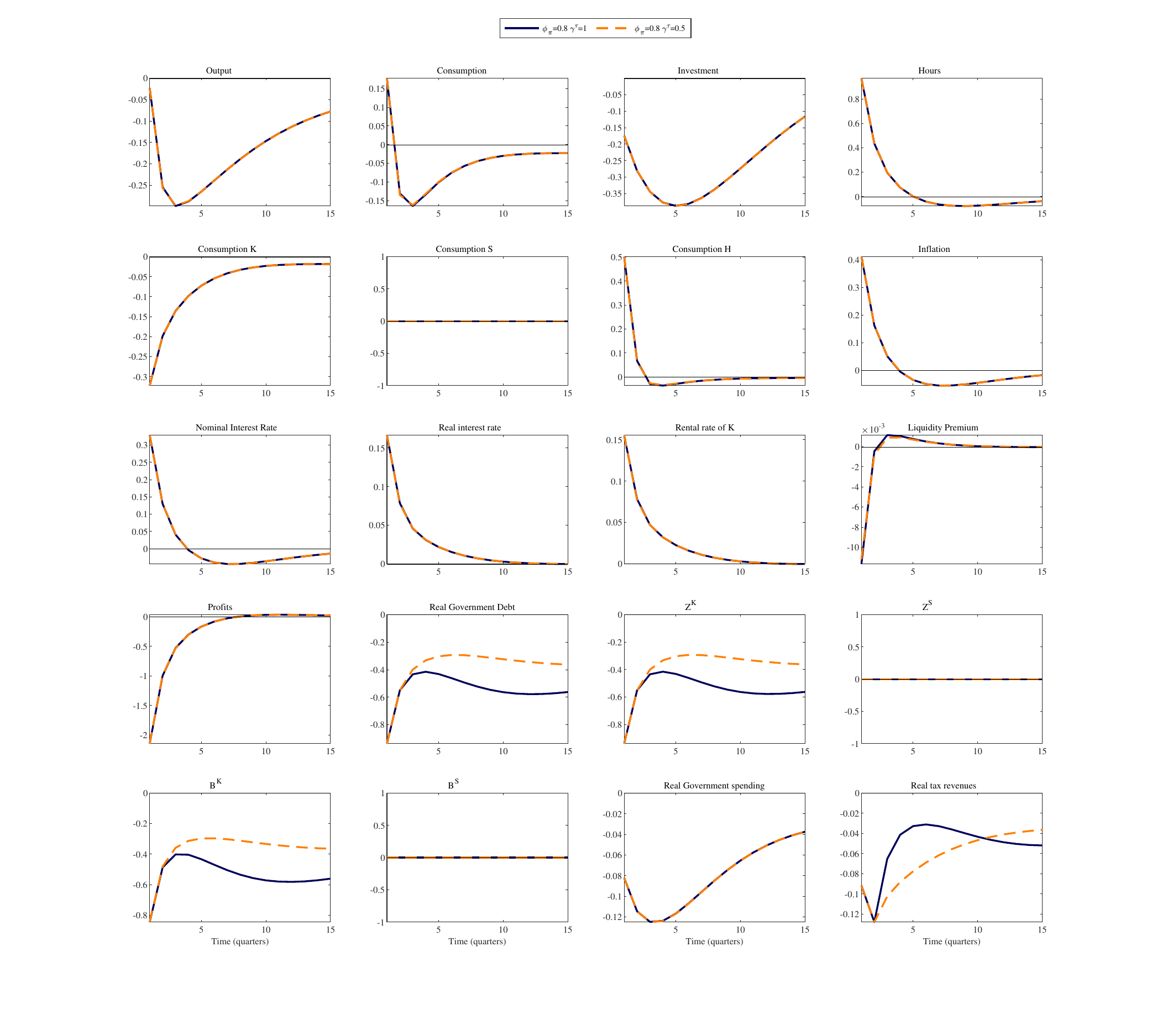}
        \caption{THANK: Impulse responses to a temporary 1\% increase in A for the two fiscal policy regimes when monetary policy is passive ($\phi_{\pi}=0.8$).\\
        \begin{tiny}
            Note: All variables are expressed in real terms except for Hours, Inflation, and Nominal interest rate. All variables related to fiscal policy are in \% deviation from the steady state of output. The remaining variables are in \% deviations from their steady state. We plot the next period realized rental rate of capital ($R^K_{t+1}$). Consumption, Z's, and B's are island-wide figures (multiplied by the population sizes $\Pi$'s).
        \end{tiny}}
        \label{fig:2A_PM}
    \end{figure}

    \begin{figure}[h!]
        \centering
        \includegraphics[width=\textwidth]{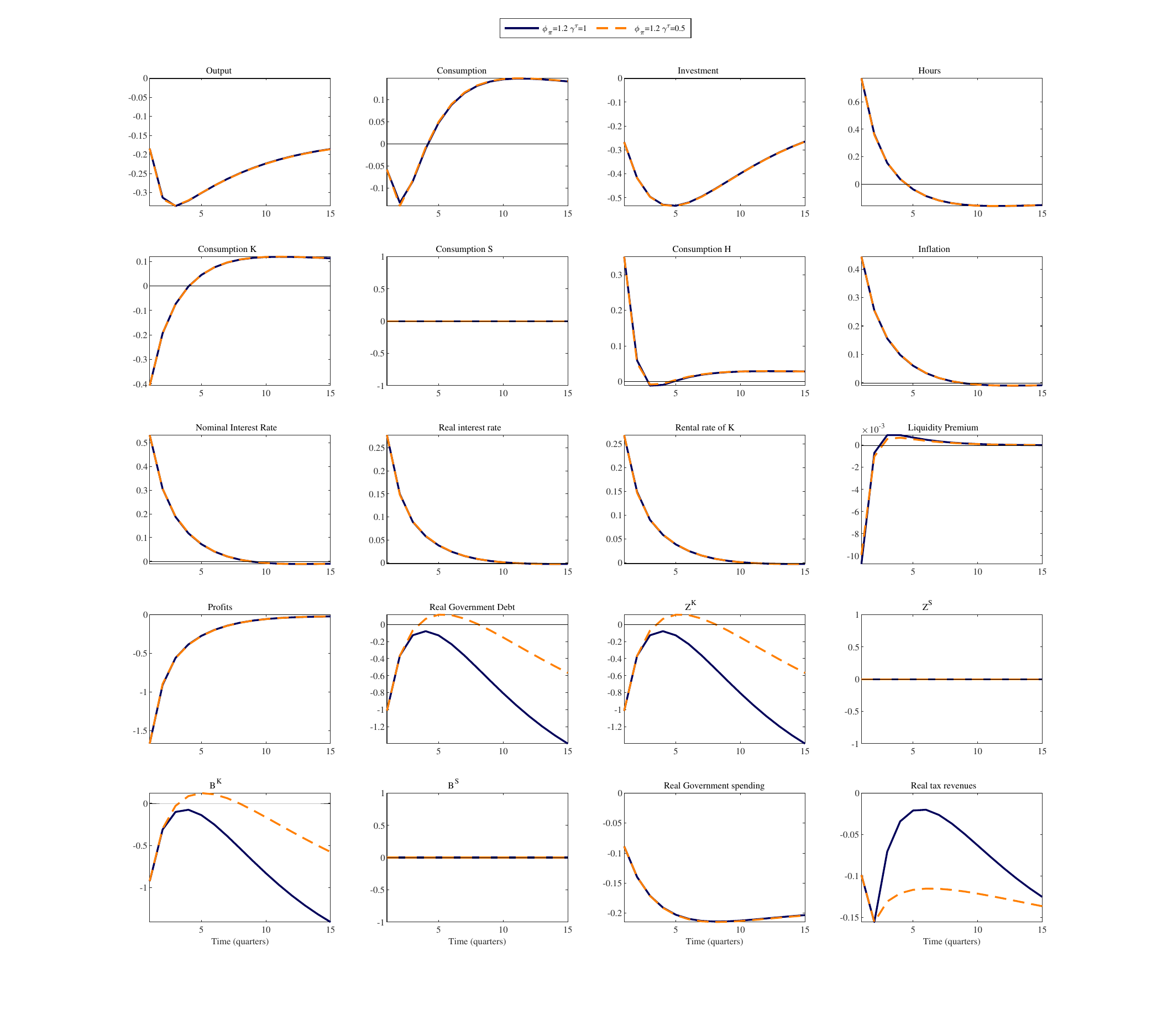}
        \caption{THANK: Impulse responses to a temporary 1\% increase in A for the two fiscal policy regimes when monetary policy is active ($\phi_{\pi}=1.2$).\\
        \begin{tiny}
            Note: All variables are expressed in real terms except for Hours, Inflation, and Nominal interest rate. All variables related to fiscal policy are in \% deviation from the steady state of output. The remaining variables are in \% deviations from their steady state. We plot the next period realized rental rate of capital ($R^K_{t+1}$). Consumption, Z's, and B's are island-wide figures (multiplied by the population sizes $\Pi$'s).
        \end{tiny}}
        \label{fig:2A_PM_Z}
        \end{figure}

    \begin{figure}[h!]
    \centering
    \includegraphics[width=\textwidth]{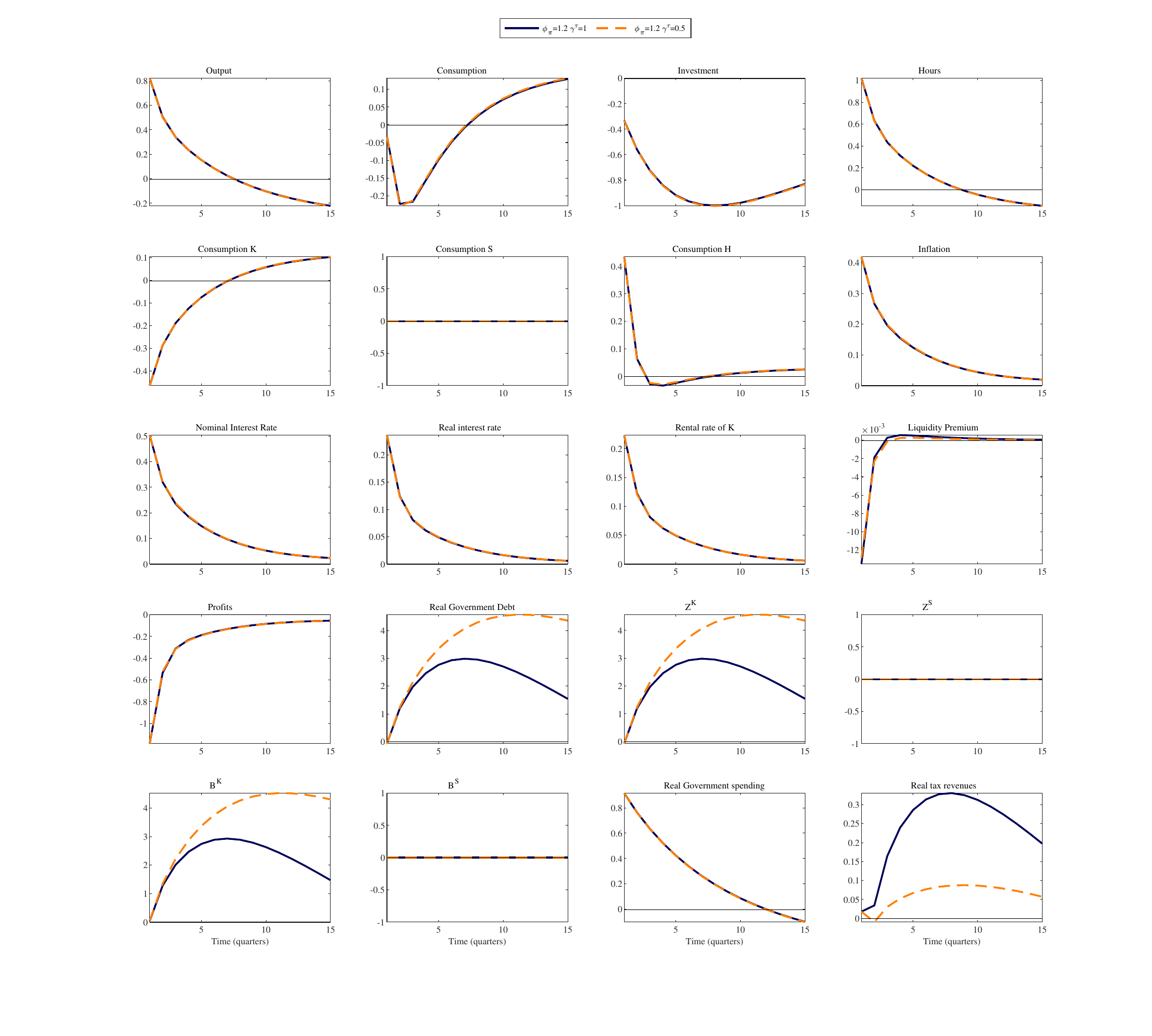}
    \caption{THANK: Impulse responses to a temporary 1\% increase in G for the two fiscal policy regimes when monetary policy is active ($\phi_{\pi}=1.2$).\\
    \begin{tiny}
        Note: All variables are expressed in real terms except for Hours, Inflation, and Nominal interest rate. All variables related to fiscal policy are in \% deviation from the steady state of output. The remaining variables are in \% deviations from their steady state. We plot the next period realized rental rate of capital ($R^K_{t+1}$). Consumption, Z's, and B's are island-wide figures (multiplied by the population sizes $\Pi$'s).
    \end{tiny}}
    \label{fig:2A_AF}
    \end{figure}

    \begin{figure}[h!]
        \centering
        \includegraphics[width=\textwidth]{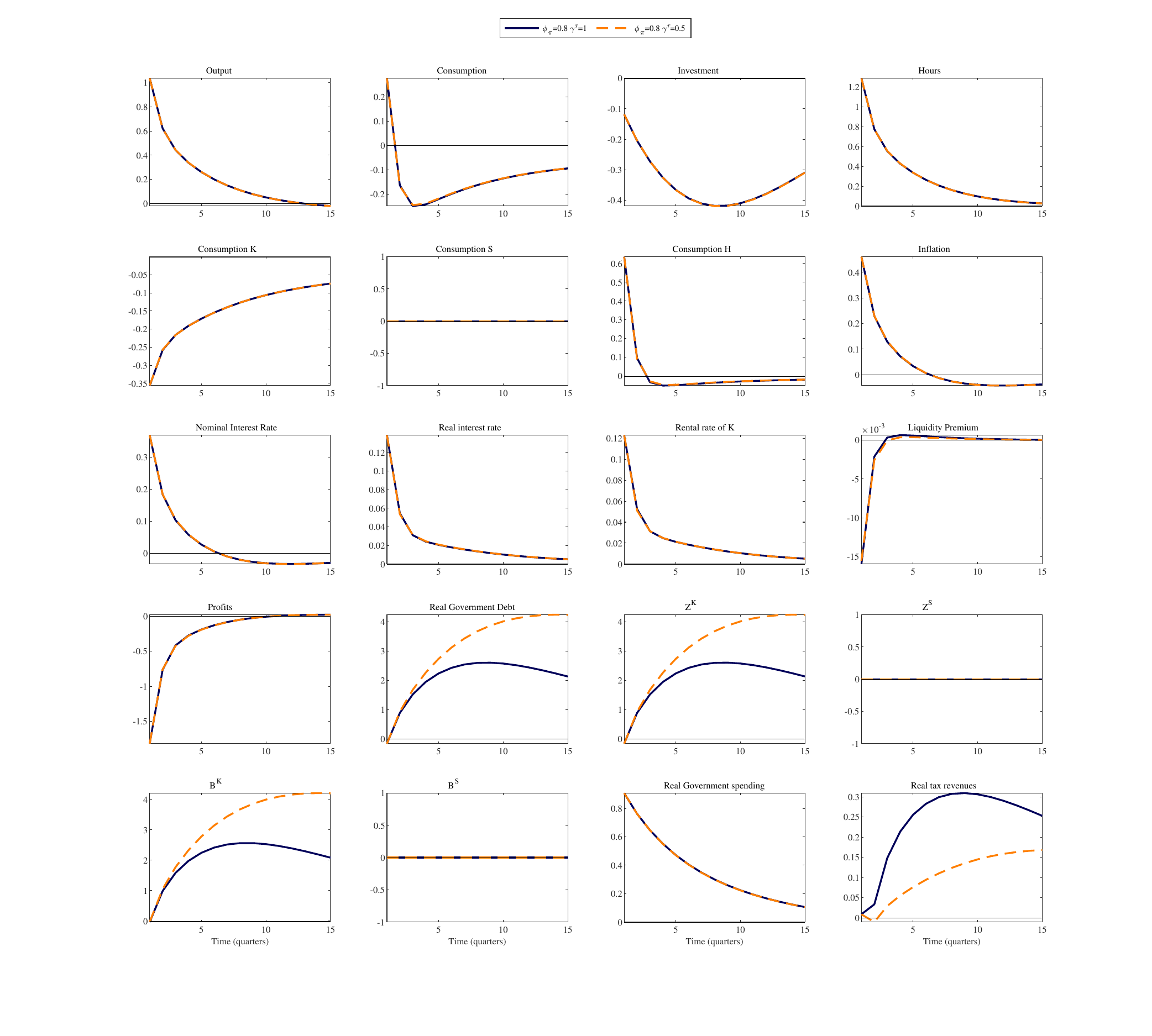}
        \caption{THANK: Impulse responses to a temporary 1\% increase in G for the two fiscal policy regimes when monetary policy is passive ($\phi_{\pi}=0.8$).\\
        \begin{tiny}
            Note: All variables are expressed in real terms except for Hours, Inflation, and Nominal interest rate. All variables related to fiscal policy are in \% deviation from the steady state of output. The remaining variables are in \% deviations from their steady state. We plot the next period realized rental rate of capital ($R^K_{t+1}$). Consumption, Z's, and B's are island-wide figures (multiplied by the population sizes $\Pi$'s).
        \end{tiny}}
        \label{fig:2A_PF}
        \end{figure}

        \begin{figure}[h!]
            \centering
            \includegraphics[width=\textwidth]{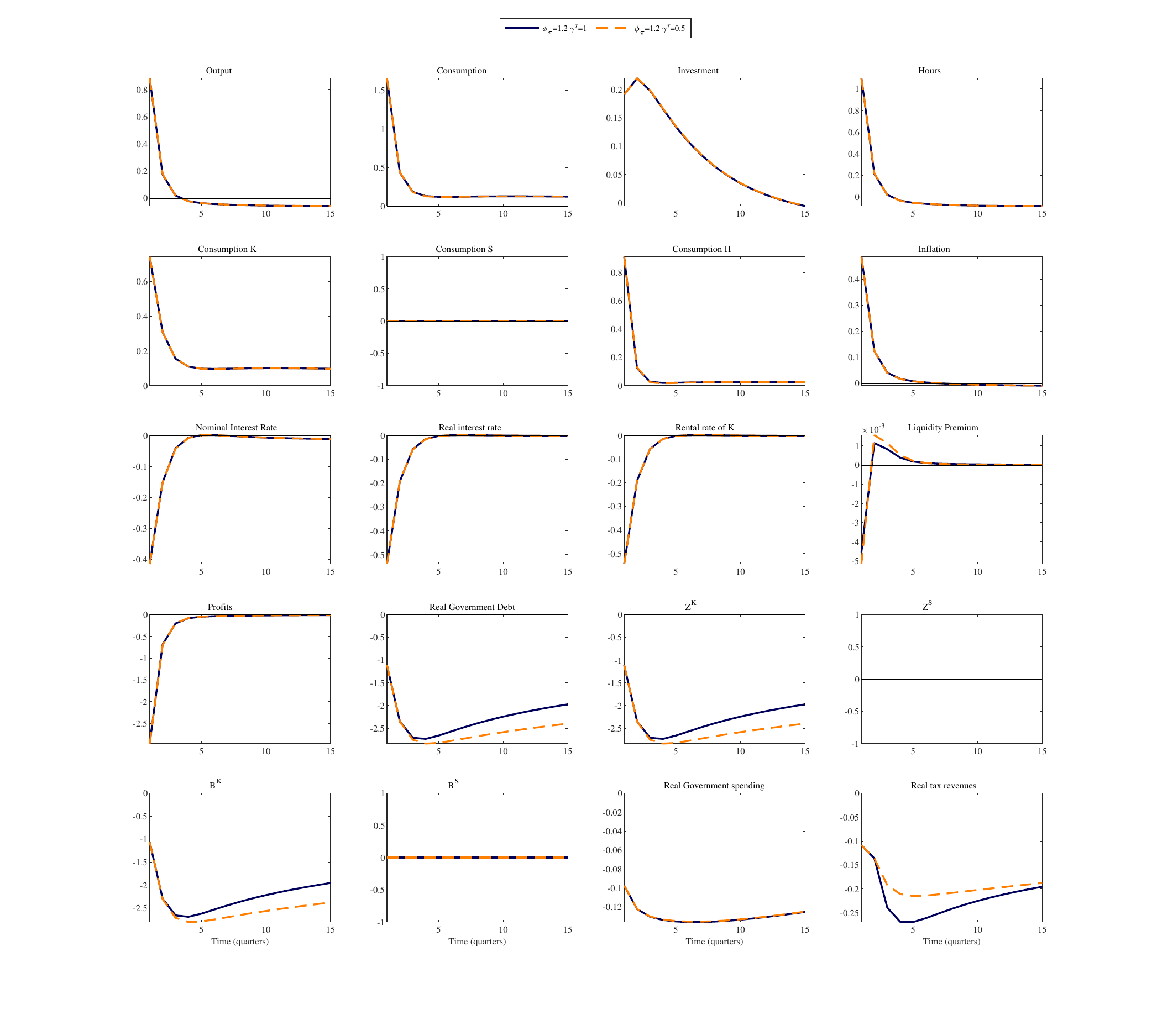}
            \caption{THANK: Impulse responses to a temporary 1\% increase in M the two fiscal policy regimes when monetary policy is active ($\phi_{\pi}=1.2$).\\
            \begin{tiny}
                Note: All variables are expressed in real terms except for Hours, Inflation, and Nominal interest rate. All variables related to fiscal policy are in \% deviation from the steady state of output. The remaining variables are in \% deviations from their steady state. We plot the next period realized rental rate of capital ($R^K_{t+1}$). Consumption, Z's, and B's are island-wide figures (multiplied by the population sizes $\Pi$'s).
            \end{tiny}}
            \label{fig:2A_AM_epsM}
        \end{figure}
    
    \begin{figure}[h!]
        \centering
        \includegraphics[width=\textwidth]{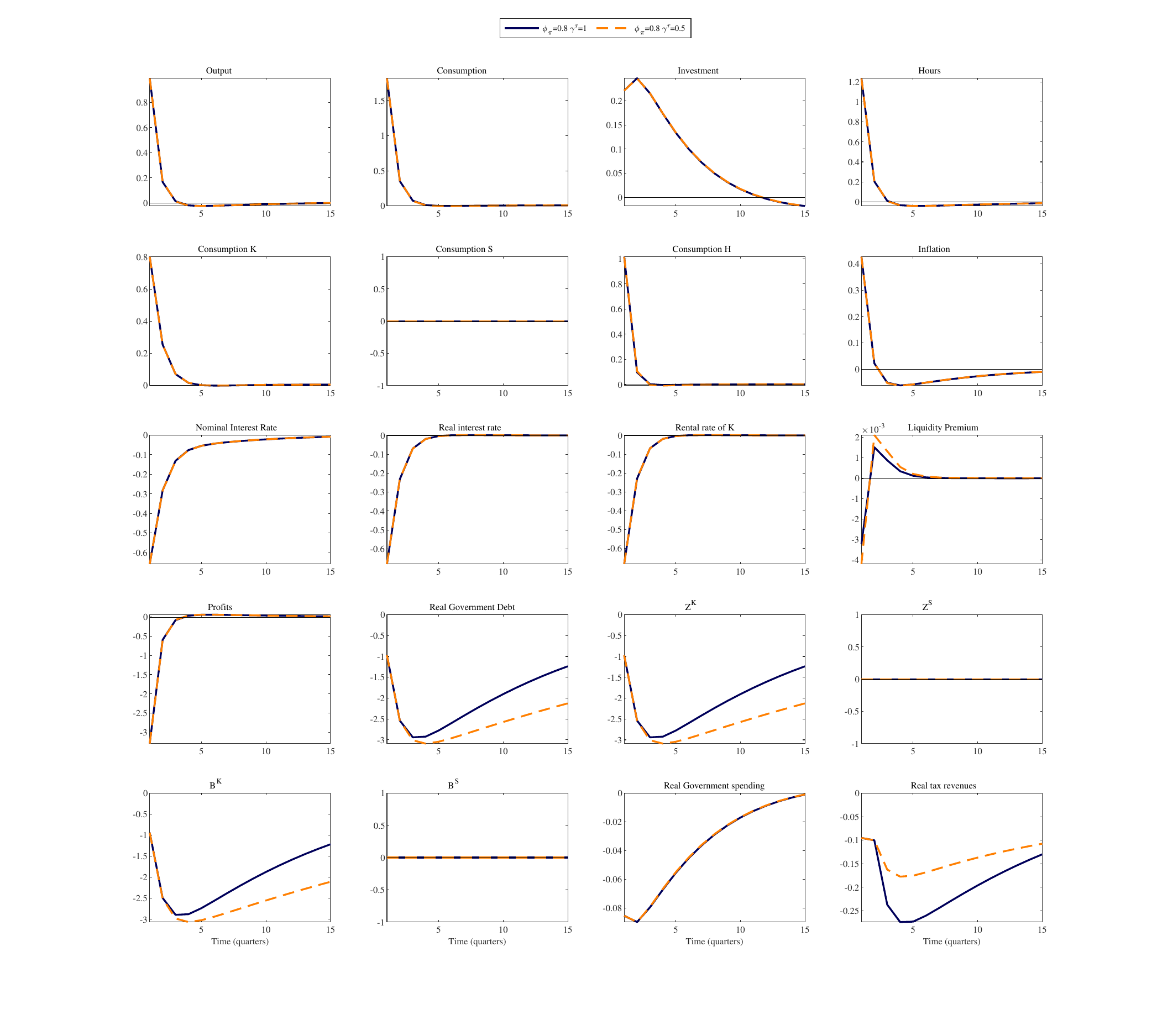}
        \caption{THANK: Impulse responses to a temporary 1\% increase in M the two fiscal policy regimes when monetary policy is passive ($\phi_{\pi}=0.8$).\\
        \begin{tiny}
            Note: All variables are expressed in real terms except for Hours, Inflation, and Nominal interest rate. All variables related to fiscal policy are in \% deviation from the steady state of output. The remaining variables are in \% deviations from their steady state. We plot the next period realized rental rate of capital ($R^K_{t+1}$). Consumption, Z's, and B's are island-wide figures (multiplied by the population sizes $\Pi$'s).
        \end{tiny}}
        \label{fig:2A_PM}
    \end{figure}

        \clearpage
        \newpage

\section{Capital portability}\label{sec:K_nodrop}

He we relax the assumption that when moving across types, households can only carry bonds with them and we allow for capital portability. As both savers and hand-to-mouth are not able to invest further in physical capital they will have to consume it all in the period they arrive to the new island. 
    As a result the modified budget constraints of the agents are as follows:
    \begin{equation}
     P_t C^{K}_{t} + Z^{K}_{t+1} + P_{t}I^K_t = P_{t}W_{t}N^{K}_{t} + (1 + R^{b}_{t-1}) \frac{\mathbb{
     B}^{K}_{t}}{\Pi_{K}} + \lambda_{K,K} R^{k}_{t}P_{t} K^S_{t} + P_t \frac{D_{t}}{\Pi_{K}} - T_t^{K}- \tau^{K}, \label{bc_k_nodrop}
    \end{equation}
    \begin{equation}
    P_t C^{S}_{t} + Z^{S}_{t+1} = P_tW_{t}N^{S}_{t} + (1 + R^{b}_{t-1}) \frac{\mathbb{B}^{S}_{t}}{\Pi_{S}}+ \lambda_{K,S} R^{k}_{t}P_{t} K^S_{t} - T_t^{S}- \tau^{S}.\label{bc_n_nodrop}
    \end{equation}
    \begin{equation}
    P_t C^{H}_{t}= P_tW_{t}N^{H}_{t} + (1+R^{b}_{t-1}) \frac{\mathbb{B}^{H}_{t}}{\Pi_{H}}+ \lambda_{K,H} R^{k}_{t}P_{t} K^S_{t} - T_t^{H}- \tau^{H}.\label{bc_u_nodrop}
    \end{equation}
    The rest of the model is the same. The results are presented in the figures below. We find that the impulse response functions are barely affected by this assumption. The reason is clear by inspecting the way capital returns enter the budget constraints of savers and hand-to-mouth consumers. The capital returns are multiplied by the transition probabilities ($\lambda_{K,S}$ and $\lambda_{K,H}$), which are very small in our calibration. Therefore, while we observe some wealth effect in comparing Figures in the main text with the Figures below, we note that these appear to be of second-order importance.

    \begin{figure}[h!]
        \centering
        \includegraphics[width=\textwidth]{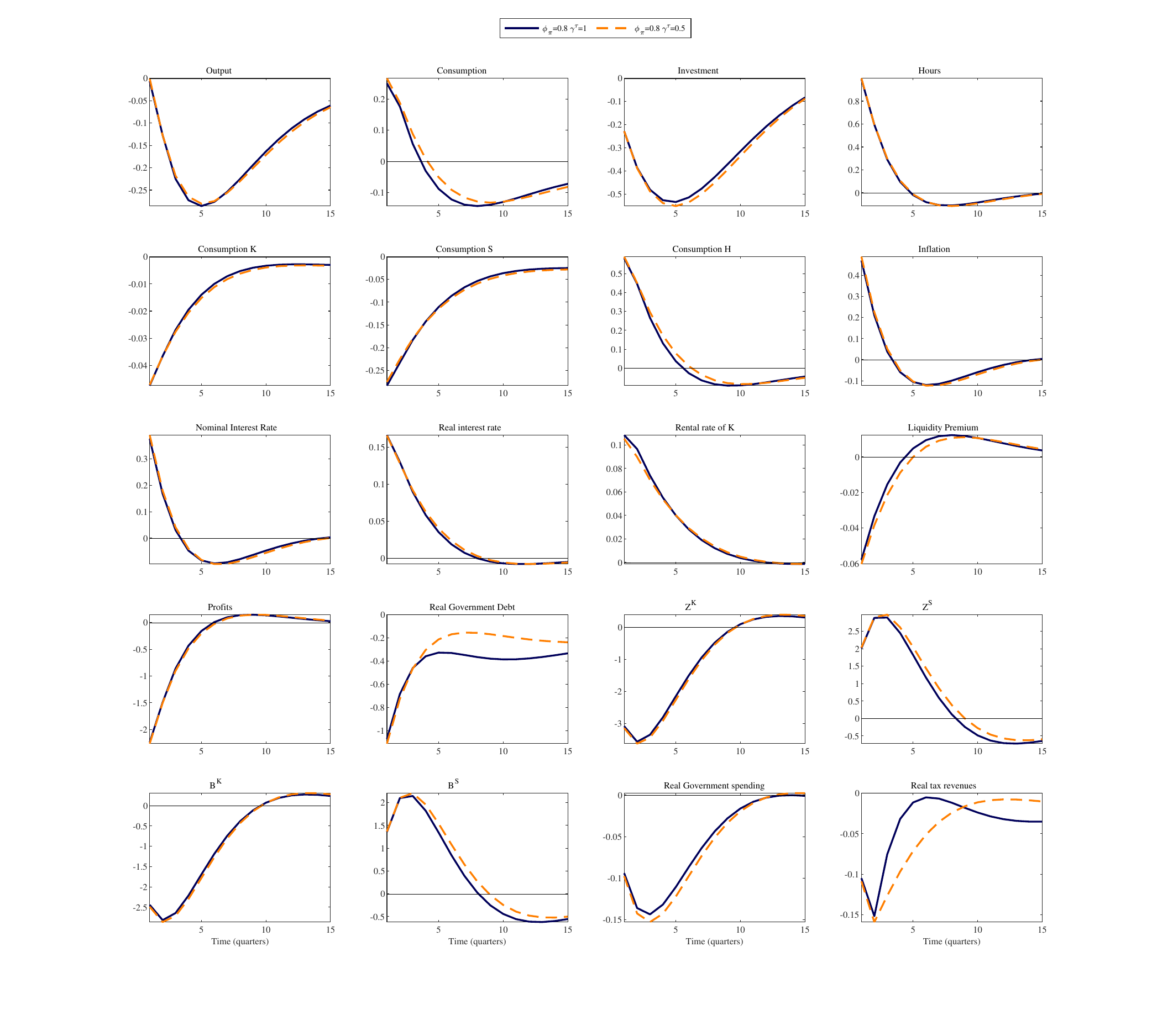}
        \caption{Impulse responses to a temporary 1\% decrease in A for the two fiscal policy regimes when monetary policy is passive ($\phi_{\pi}=0.8$) in the model with capital portability.\\
        \begin{tiny}
            Note: All variables are expressed in real terms except for Hours, Inflation, and Nominal interest rate. All variables related to fiscal policy are in \% deviation from the steady state of output. The remaining variables are in \% deviations from their steady state. We plot the next period realized rental rate of capital ($R^K_{t+1}$). Consumption, Z's, and B's are island-wide figures (multiplied by the population sizes $\Pi$'s).
        \end{tiny}}
        \label{fig:3A_PM_epsZ_noK}
    \end{figure}
    
\begin{figure}[h!]
        \centering
        \includegraphics[width=\textwidth]{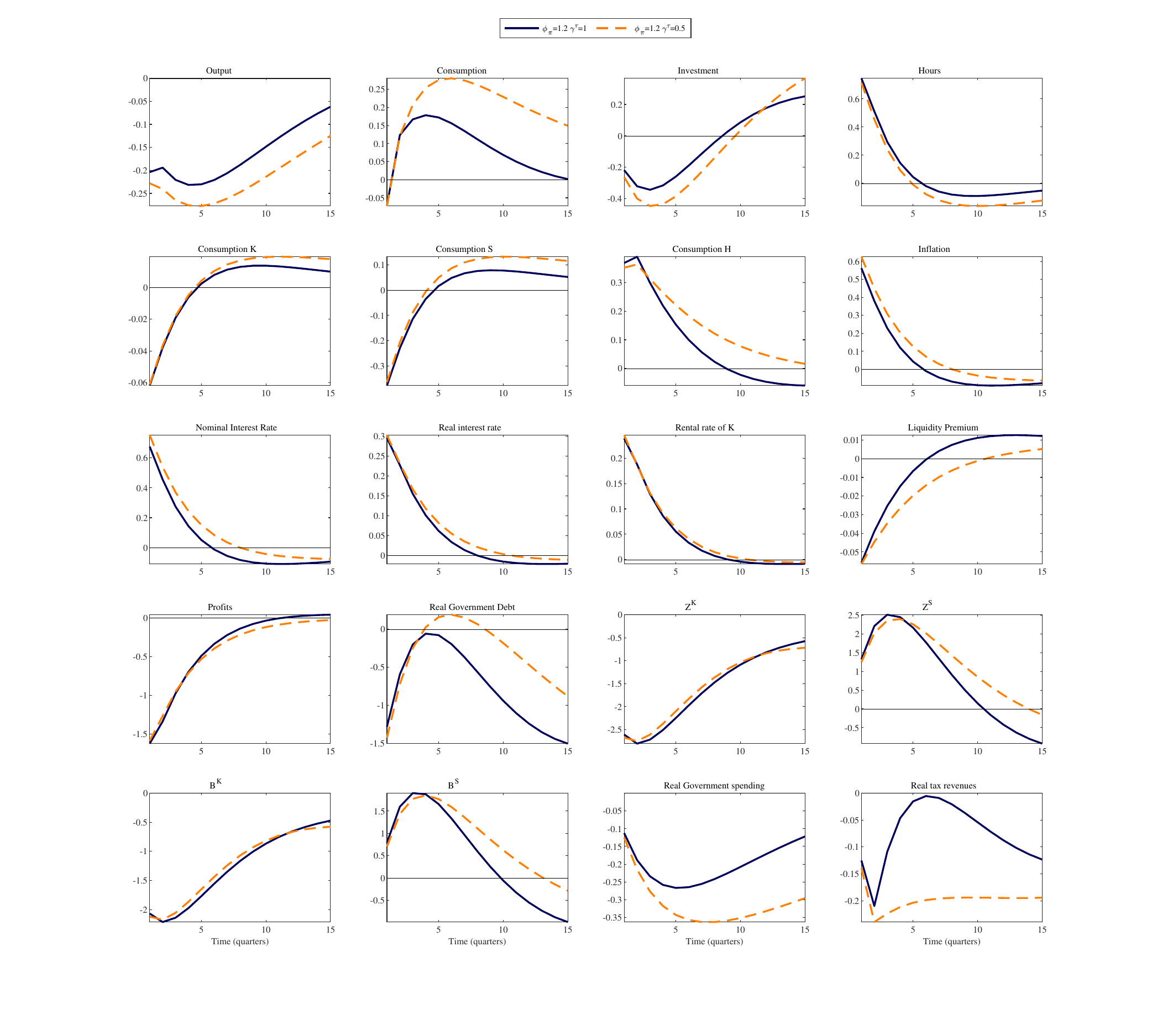}
        \caption{Impulse responses to a temporary 1\% decrease in A for the two fiscal policy regimes when monetary policy is active ($\phi_{\pi}=1.2$) in the model with capital portability.\\
        \begin{tiny}
            Note: All variables are expressed in real terms except for Hours, Inflation, and Nominal interest rate. All variables related to fiscal policy are in \% deviation from the steady state of output. The remaining variables are in \% deviations from their steady state. We plot the next period realized rental rate of capital ($R^K_{t+1}$). Consumption, Z's, and B's are island-wide figures (multiplied by the population sizes $\Pi$'s).
        \end{tiny}}
        \label{fig:3A_AM_epsZ_noK}
    \end{figure}  

    \begin{figure}[h!]
        \centering
        \includegraphics[width=\textwidth]{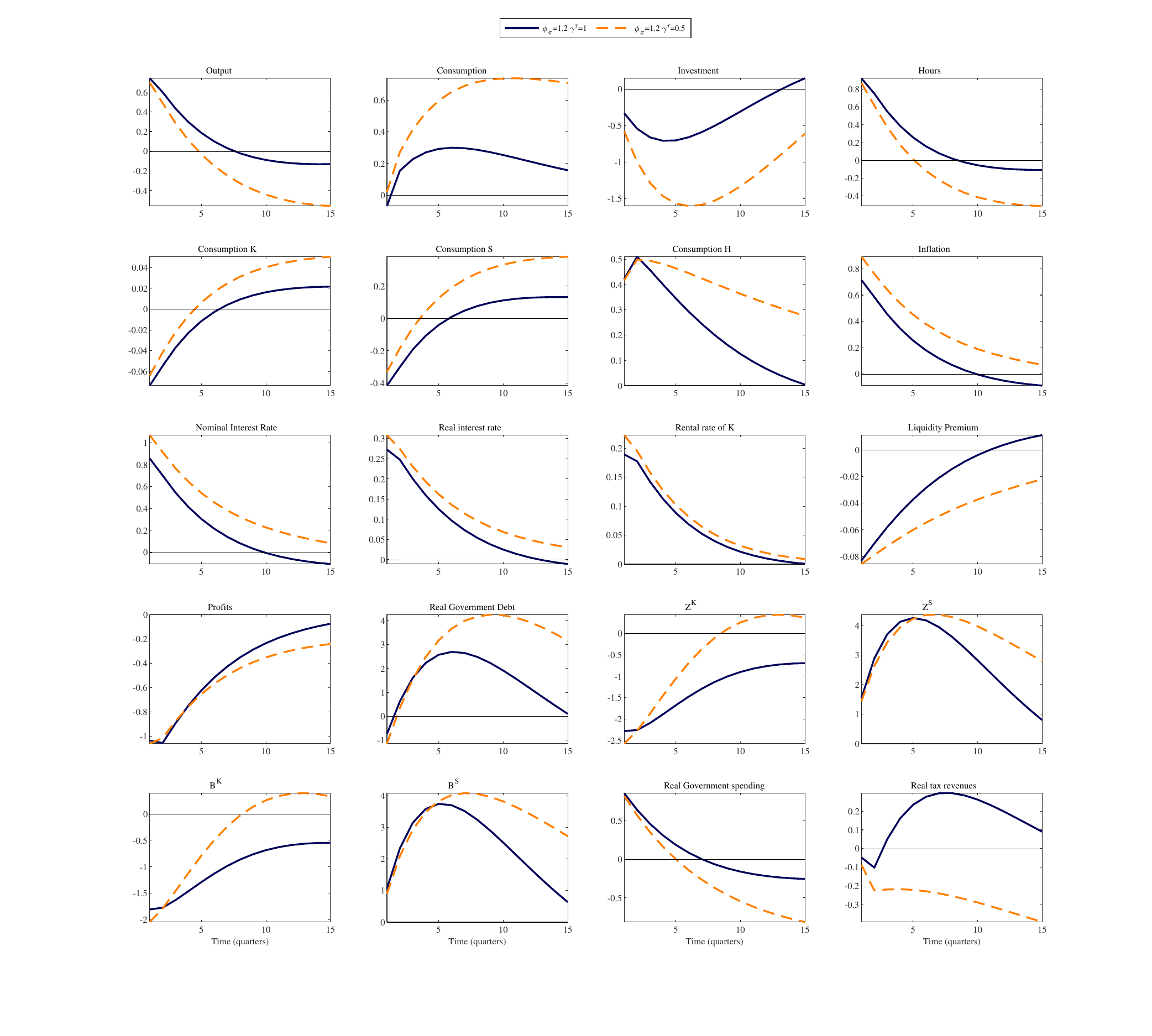}
        \caption{Impulse responses to a temporary 1\% increase in $G^N$ for the two fiscal policy regimes when monetary policy is active ($\phi_{\pi}=1.2$) in the model with capital portability.\\
        \begin{tiny}
            Note: All variables are expressed in real terms except for Hours, Inflation, and Nominal interest rate. All variables related to fiscal policy are in \% deviation from the steady state of output. The remaining variables are in \% deviations from their steady state. We plot the next period realized rental rate of capital ($R^K_{t+1}$). Consumption, Z's, and B's are island-wide figures (multiplied by the population sizes $\Pi$'s).
        \end{tiny}}
        \label{fig:3A_AM_noK}
        \end{figure}
    
        \begin{figure}[h!]
            \centering
            \includegraphics[width=\textwidth]{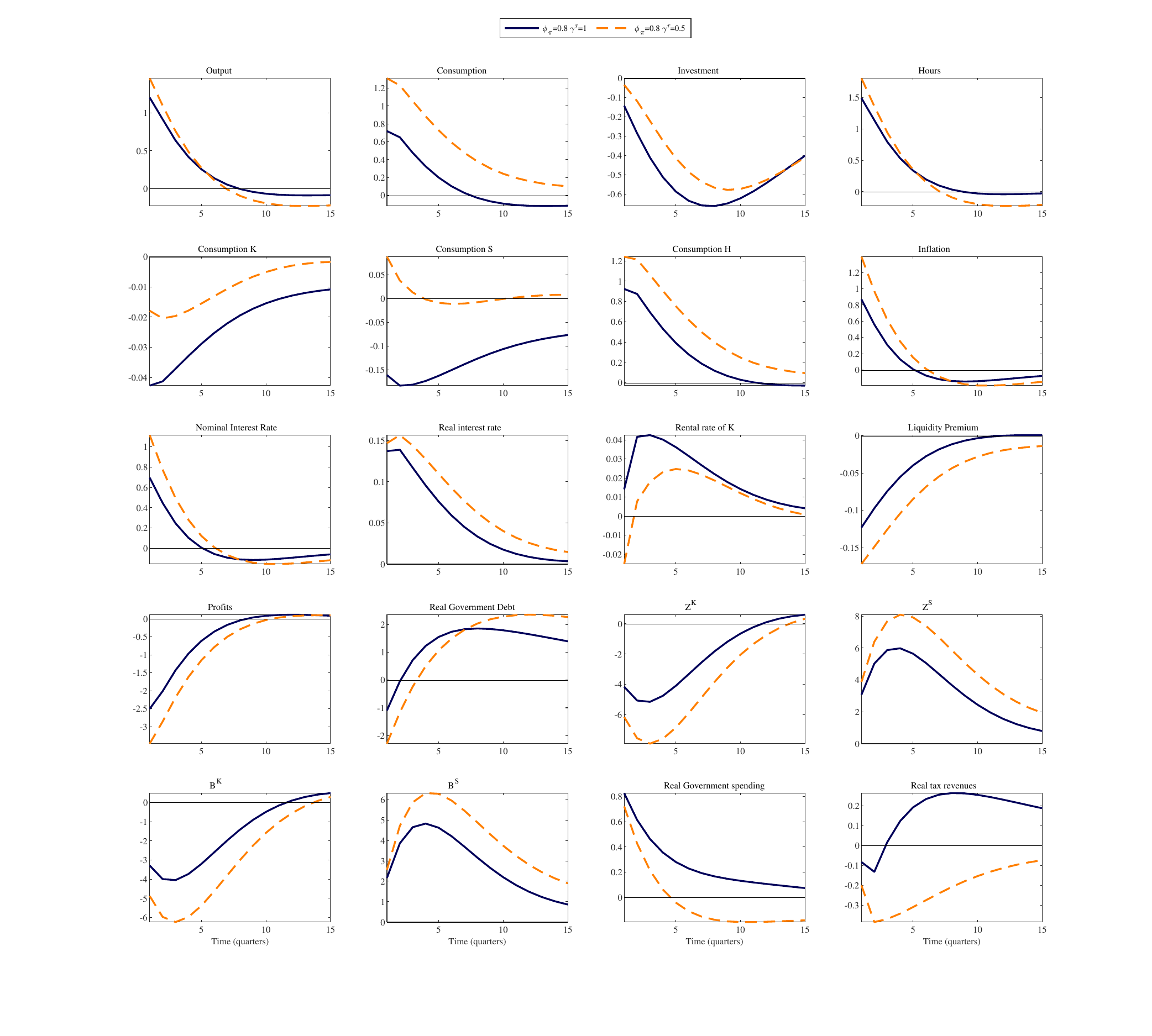}
            \caption{Impulse responses to a temporary 1\% increase in $G^N$ ffor the two fiscal policy regimes when monetary policy is passive ($\phi_{\pi}=0.8$) in the model with capital portability.\\
              \begin{tiny}
                Note: All variables are expressed in real terms except for Hours, Inflation, and Nominal interest rate. All variables related to fiscal policy are in \% deviation from the steady state of output. The remaining variables are in \% deviations from their steady state. We plot the next period realized rental rate of capital ($R^K_{t+1}$). Consumption, Z's, and B's are island-wide figures (multiplied by the population sizes $\Pi$'s).
        \end{tiny}}
            \label{fig:3A_PM_noK}
        \end{figure}

        \begin{figure}[h!]
            \centering
            \includegraphics[width=\textwidth]{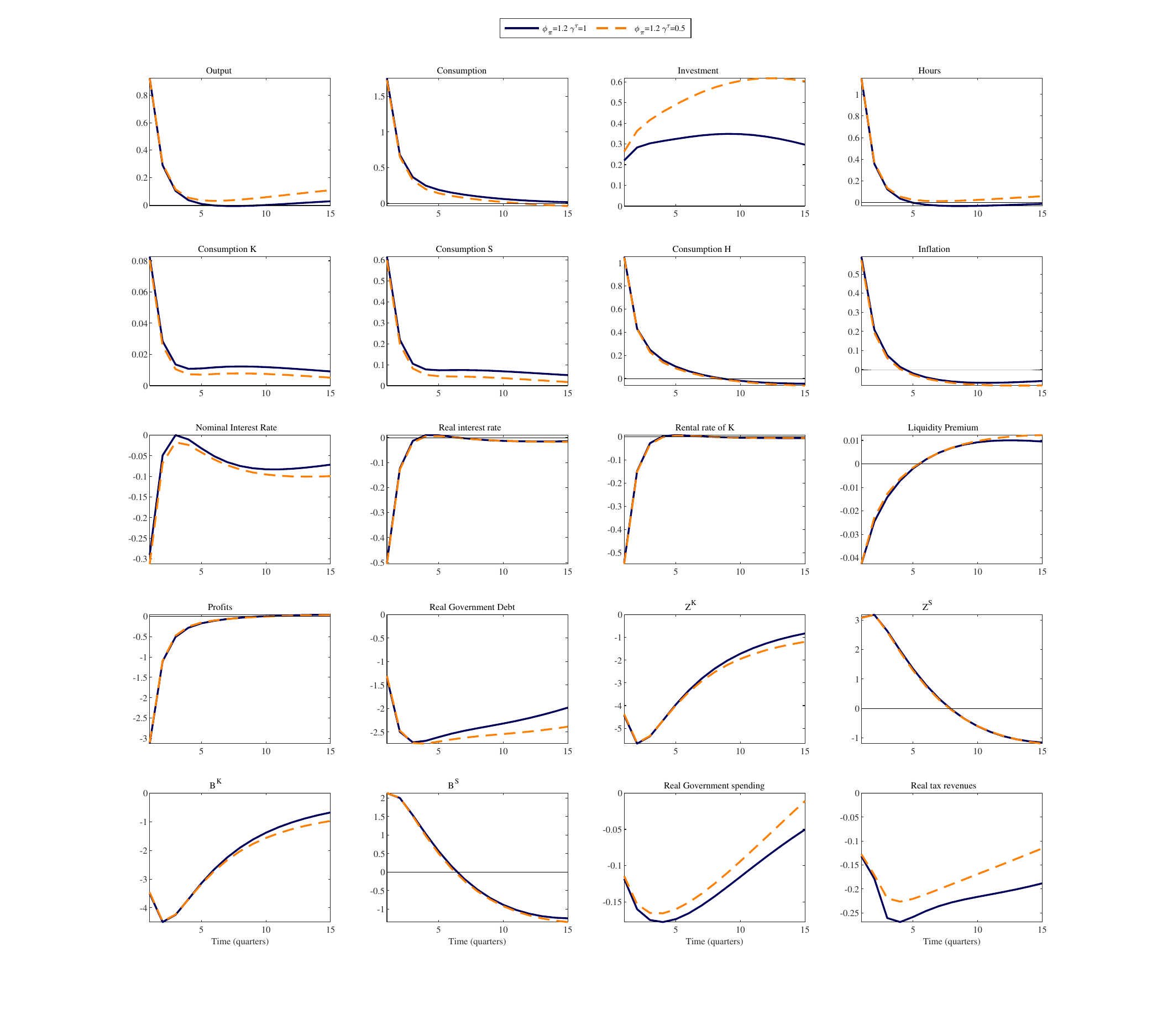}
            \caption{Impulse responses to a temporary 1\% increase in M for the two fiscal policy regimes when monetary policy is active ($\phi_{\pi}=1.2$) in the model with capital portability.\\
            \begin{tiny}
                Note: All variables are expressed in real terms except for Hours, Inflation, and Nominal interest rate. All variables related to fiscal policy are in \% deviation from the steady state of output. The remaining variables are in \% deviations from their steady state. We plot the next period realized rental rate of capital ($R^K_{t+1}$). Consumption, Z's, and B's are island-wide figures (multiplied by the population sizes $\Pi$'s).
            \end{tiny}}
            \label{fig:3A_AM_epsM_noK}
            \end{figure}
            
            \begin{figure}[h!]
                \centering
                \includegraphics[width=\textwidth]{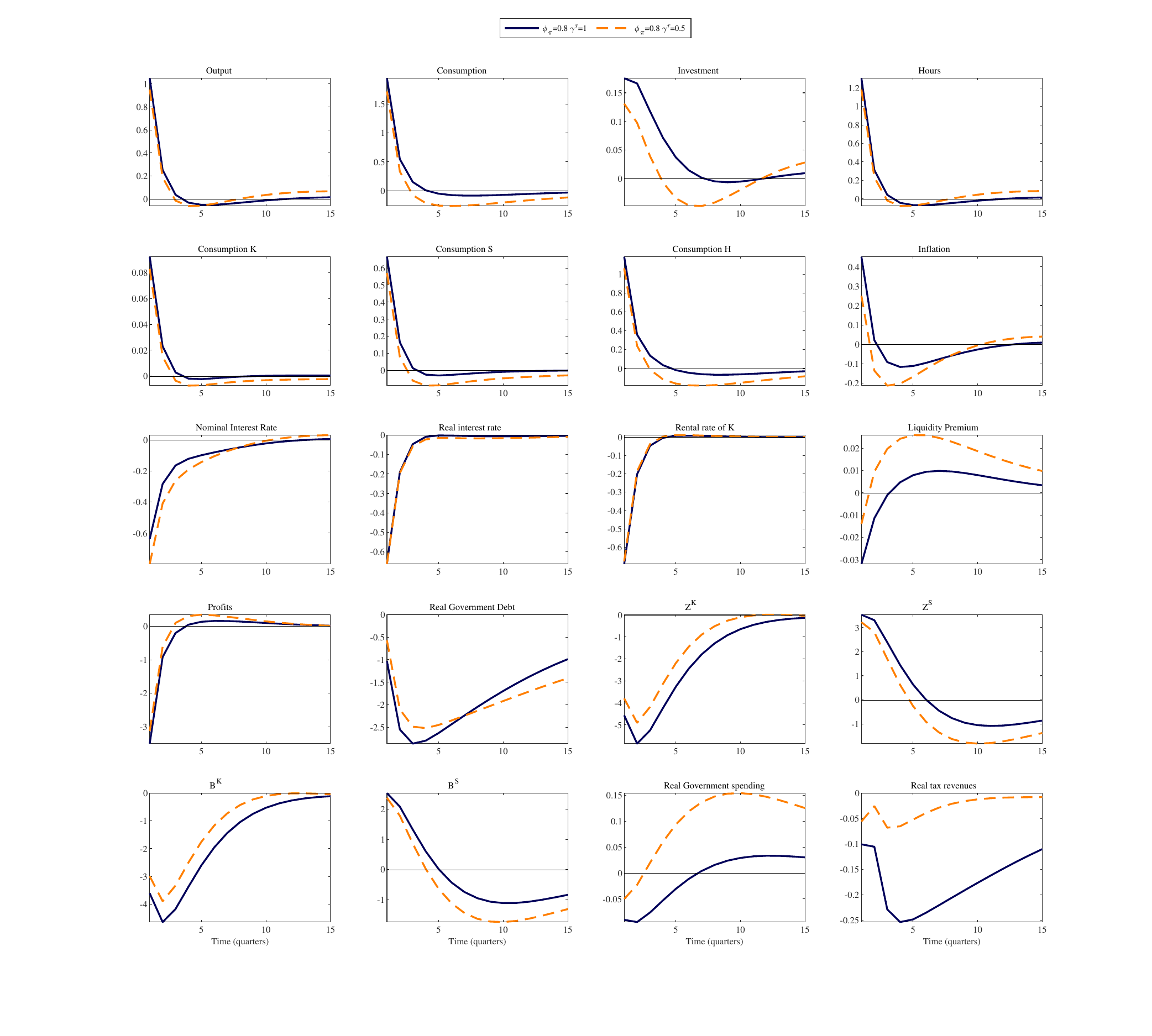}
                \caption{Impulse responses to a temporary 1\% increase in M for the two fiscal policy regimes when monetary policy is passive ($\phi_{\pi}=0.8$) in the model with capital portability.\\
                    \begin{tiny}
                    Note: All variables are expressed in real terms except for Hours, Inflation, and Nominal interest rate. All variables related to fiscal policy are in \% deviation from the steady state of output. The remaining variables are in \% deviations from their steady state. We plot the next period realized rental rate of capital ($R^K_{t+1}$). Consumption, Z's, and B's are island-wide figures (multiplied by the population sizes $\Pi$'s).
            \end{tiny}}
                \label{fig:3A_PM_epsM_noK}
            \end{figure}

\end{document}